\documentclass{agujournal}

\journalname{Planetary and Space Science}

\usepackage{xcolor} 
\usepackage[normalem]{ulem} 

\usepackage[colorinlistoftodos]{todonotes}

\begin{document}

%
%


\title{Study of gravity waves distribution and propagation in the thermosphere of Mars based on MGS, ODY, MRO and MAVEN density measurements}

%
%




\authors{M. Vals\affil{1}, A. Spiga\affil{1,2}, F. Forget\affil{1}, E. Millour\affil{1}, L. Montabone\affil{3}, F. Lott\affil{1}}


\affiliation{1}{Laboratoire de M\'{e}t\'{e}orologie Dynamique (LMD/IPSL), Sorbonne Universit\'{e}, Centre National de la Recherche Scientifique, {\'E}cole Normale Sup{\'e}rieure, {\'E}cole Polytechnique, Paris, France}
\affiliation{2}{Institut Universitaire de France, Paris, France}
\affiliation{3}{Space Science Institute, Greenbelt, MD, USA}




\correspondingauthor{M. Vals}{margaux.vals@lmd.jussieu.fr}




\begin{keypoints}
\item Gravity wave activity causes density perturbations in the Martian thermosphere.
\item MAVEN found a correlation between gravity wave activity and inverse background temperature.
\item Lower-altitude aerobraking measurements do not show this correlation, except for Mars Odyssey.
\item Aerobraking data and climate models suggest instead wave activity correlated with static stability.
\item No such correlation in low latitudes points to a mix of saturation, critical levels and sources.
\end{keypoints}

%
%


\begin{abstract}
\emph{By  measuring  the regular  oscillations  of  the  density  of CO$_{2}$
in  the  upper  atmosphere (between  120  and
190~km), the mass spectrometer MAVEN/NGIMS 
(Atmosphere  and  Volatile  EvolutioN/Neutral  Gas  Ion
Mass  Spectrometer) reveals  the local  impact of gravity waves.
This yields precious information on
the activity of gravity waves
and the atmospheric conditions in which
they propagate and break.
The intensity of gravity waves 
measured by MAVEN in the upper atmosphere
has been shown to be dictated 
by saturation processes in isothermal conditions.
As a result, gravity waves activity is correlated to the evolution of the inverse of the 
background temperature.
Previous data gathered at lower altitudes ($\sim$95 to $\sim$150~km) 
during aerobraking by the accelerometers 
on board MGS (Mars Global Surveyor), ODY (Mars Odyssey) 
and MRO (Mars Reconnaissance Orbiter)
are analyzed in the light of those recent 
findings with MAVEN. 
The anti-correlation between 
GW-induced density perturbations 
and background temperature is 
plausibly found in the ODY data acquired 
in the polar regions, but not in the MGS and MRO data. 
MRO data in polar regions exhibit a correlation between 
the density perturbations and 
the Brunt-V\"{a}is\"{a}l\"{a} frequency
(or, equivalently, static stability),
obtained from Global Climate Modeling
compiled in the Mars Climate Database. 
At lower altitude levels (between  100  and 120~km), 
although wave saturation might still be dominant, 
isothermal conditions are no longer verified. In this case, 
theory predicts that the intensity 
of gravity waves is no more correlated to background 
temperature, but to static stability. At other 
latitudes in the three aerobraking datasets,
the GW-induced relative density perturbations
are correlated with 
neither inverse temperature
nor static stability;
in this particular case, 
this means that
the observed activity of gravity waves
is not only controlled by saturation,
but also by the effects of gravity-wave sources 
and wind filtering through critical levels. 
This result highlights 
the exceptional nature of MAVEN/NGIMS observations 
which combine both isothermal and saturated conditions
contrary to aerobraking measurements.}
\end{abstract}

%
%

%


%
%
%
%

\section{Introduction}

Gravity waves propagate as perturbations of the stratified atmospheric fluid \citep{Goss:75}, 
with the buoyancy force being the restoring mechanism 
giving rise to the waves \citep[cf][for a review]{Frit:03,Alex:10}. 
While being essentially regional-scale phenomena, 
gravity waves can be responsible for 
significant dynamical and thermal forcing of the global atmospheric state, 
as they transfer their momentum and energy 
upon their saturation and breaking in the upper atmosphere \citep{Lind:81,Palm:86,McFa:87}. 

Gravity waves are ubiquitous in the Martian atmosphere 
and were actually one of the first atmospheric phenomenon
to be witnessed by orbiting spacecraft \citep{Brig:74}. 
As is the case on Earth \citep{Osul:95,Vinc:00,Plou:03,Spig:08gw}, 
those waves may be triggered in the Martian lower atmosphere 
by different sources: 
topography \citep{Pick:79,Pick:81}, 
convection \citep{Spig:13rocket,Imam:16}, or 
jet-streams and fronts in ageostrophic evolution.
Amongst all those sources, 
only the impact of the topographic source on the global circulation is accounted for 
in all Martian Global Climate Models \citep[GCM, e.g.][]{Barn:90,Coll:97,Forg:99,Hart:05}, 
although the exploration of the impact of an additional non-orographic source 
is a topic of current active research \citep{Medv:15,Gill:18submitted}.

The upward propagation of gravity waves 
from their tropospheric sources to the upper atmosphere 
leads to large departures of density, temperature and winds in the thermosphere, 
owing to the exponential increase of gravity wave amplitude with height \citep{Frit:03,Pari:09}. 
Measurements of CO$_{2}$ density through accelerometers, 
gathered during the aerobraking of 
Mars Global Surveyor (MGS), 
Mars Odyssey (ODY) 
and Mars Reconnaissance Orbiter (MRO) 
observed the sustained gravity wave activity 
in the Martian thermosphere between 90 and 130~km \citep{Frit:06,Crea:06ugw,Tols:07}. 
Those measurements also demonstrated 
the large variability of the gravity-wave amplitudes 
with season, local time, latitude and longitude.  

The Mars Atmosphere and Volatile Evolution (MAVEN) 
mission to Mars \citep{Jako:15maven}, 
operating since 2014, 
is dedicated to studying the upper atmosphere of Mars and, as such, 
is a unique opportunity to broaden the knowledge of gravity wave activity on Mars. 
The mass spectrometer NGIMS (Neutral Gas Ion Mass Spectrometer) on board MAVEN \citep{Maha:15} 
recently delivered new and more accurate measurements of density fluctuations 
at upper altitudes between 120 and 300~km, 
identified as typical gravity-wave signatures \citep{Yigi:15,Engl:17}. 

Based on those MAVEN/NGIMS measurements, 
\citet{Tera:17} observed that 
gravity-wave amplitudes derived from Ar density with wavelengths between $\sim$100 and $\sim$500~km near the exobase in the Martian thermosphere 
are anti-correlated with the background temperature. The authors demonstrated this anti-correlation by considering gravity waves saturation caused by convective instability in the upper thermosphere. These observations were further discussed in a recent study focusing on Ar density between 120 and 200~km by \citet{Sidd:19}, who observed that gravity waves amplitudes also increase with increasing solar zenith angle.

The goal of this paper is 
to build on those recent findings by MAVEN and 
to expand this analysis 
by comparing all available aerobraking data from other orbiting spacecraft. In particular, we explore the saturation conditions of gravity waves in a lower part of the thermosphere (between 90 and 130~km) than the one observed by MAVEN (between 120 and 300~km). 
Thus, we obtain a broader dataset of the variability of 
gravity wave activity with altitude, latitude and season.
This allows us to compare the available measurements 
with diagnostics obtained by GCM
through the Mars Climate Database \citep[MCD][]{Lewi:99,Forg:99,Mill:15}

This paper is organized as follows. 
In section~\ref{sec:data},
we provide information on the datasets.
Section~\ref{sec:maven} features a discussion
of the MAVEN/NGIMS measurements, while
section~\ref{sec:aero} features a comparative discussion
of the aerobraking datasets.
We conclude in section~\ref{sec:conclu}.

\section{Data and Method \label{sec:data}}

\subsection{Datasets used in this study}

\begin{figure*}[!ht]
 \begin{center}
\includegraphics[width=\textwidth]{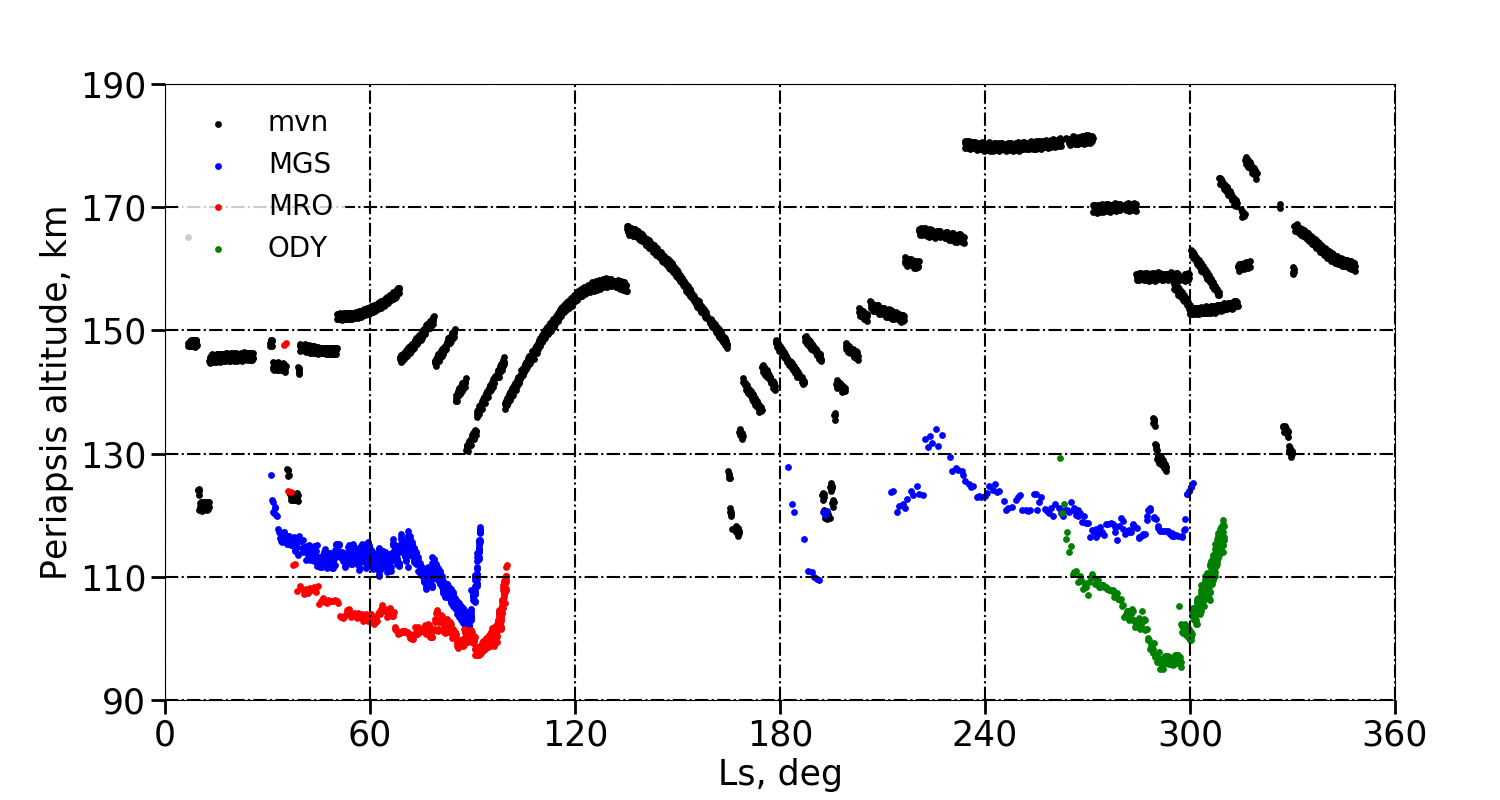}
\end{center}
\caption{Vertical (km) and seasonal (Solar Longitude in degrees) coverage of Mars Global Surveyor (MGS), Mars Odyssey (ODY), Mars Reconnaissance Orbiter (MRO) and MAVEN (MVN) spacecrafts, each dot corresponds to the periapsis location of one orbit}
\label{all_alt_Ls}
\end{figure*}

\begin{figure*}[!ht]
 \begin{center}
\includegraphics[width=\textwidth]{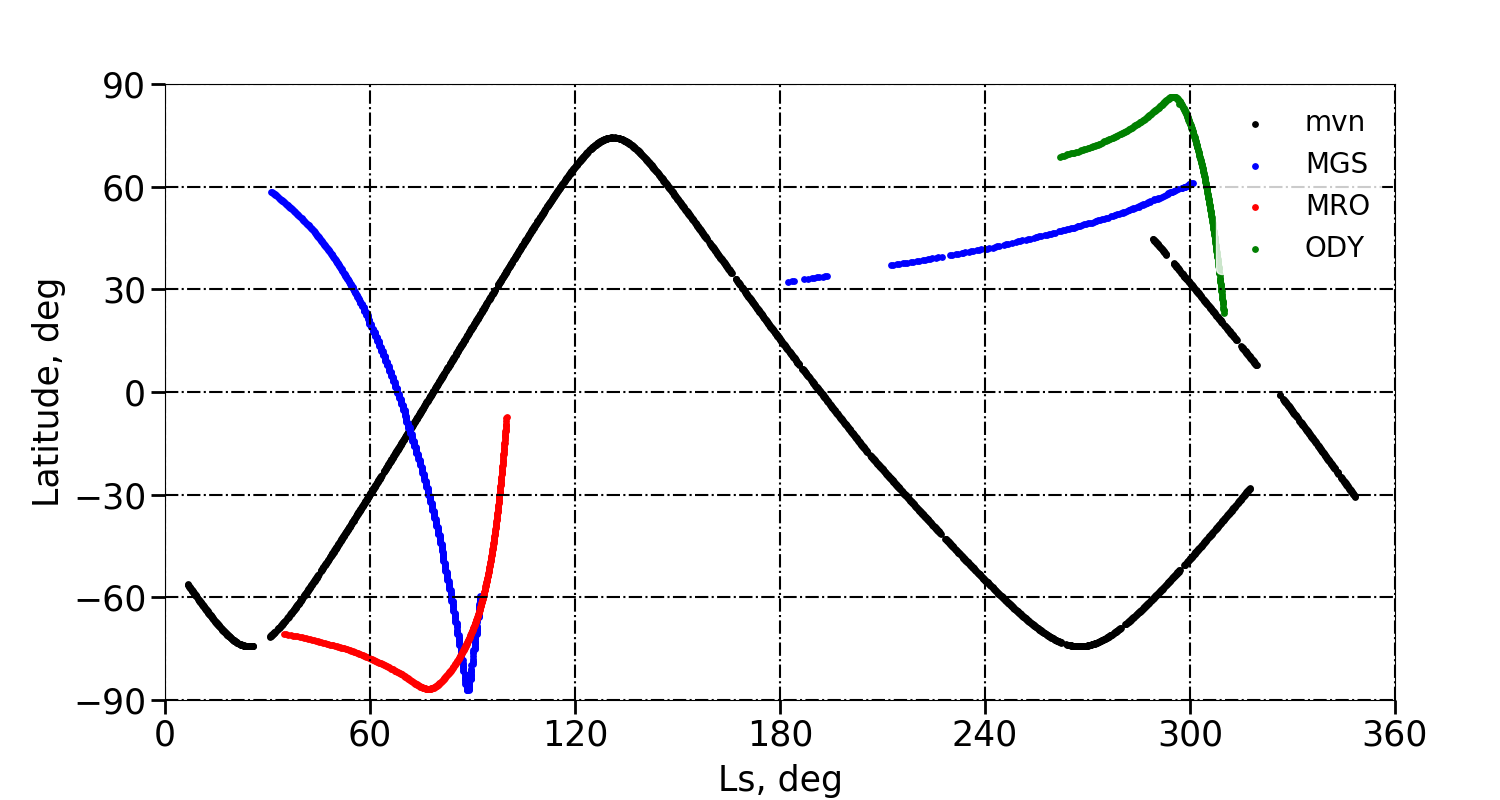}
\end{center}
\caption{Latitudinal (degrees) and seasonal (Solar Longitude in degrees) coverage of Mars Global Surveyor (MGS), Mars Odyssey (ODY), Mars Reconnaissance Orbiter (MRO) and MAVEN (MVN) spacecrafts, each dot corresponds to the periapsis location of one orbit}
\label{all_lat_Ls}
\end{figure*}

During aerobraking operations in the Martian thermosphere, 
the accelerometers of MGS, ODY and MRO \citep{Lyon:99,Smit:05,Tols:08}
acquired data during
850 passes for MRO (since September 1997, Martian Year [MY] 23) \citep{Keat:02}, 
320 passes for ODY (since October 2001, MY 25) \citep{Tols:07ody}, and 
430 passes for MGS (from April to August 2006, MY 28) \citep{Tols:07mro}, 
covering latitude ranges from 
60$^{\circ}$N to 90$^{\circ}$S for MGS, 
30$^{\circ}$N to 90$^{\circ}$N for ODY, 
and 
0$^{\circ}$ to 90$^{\circ}$S for MRO. 
Periapsis altitudes varied from about 95~km to 150~km (see Figures \ref{all_alt_Ls} and \ref{all_lat_Ls}).

In addition to those aerobraking datasets, 
the CO$_{2}$ density variations from 3124 orbits are available 
from MAVEN/NGIMS mass spectrometer data
reported in the NASA Planetary Data System 
from October 2014 (MY32) to February 2017 (MY33) \citep{Benn:14}.
Considering the NGIMS settings were changed to a new operating mode starting from February 2015, as mentioned in \citet{Engl:17} and \citet{Tera:17}, we chose to focus on datasets from February 2015 to February 2017.
The instrument is still in operation at the time of writing
and the present study can be complemented in the future 
by an analysis of the interannual variability.
The MAVEN observations cover (high periapsis) altitude ranges between $\sim$120~km and $\sim$190~km, and have large latitudinal coverage, as shown in Figure \ref{all_alt_Ls} and Figure \ref{all_lat_Ls}.



\subsection{Computing the amplitude of gravity wave perturbations}

\begin{figure*}[!tb]
\begin{center}
\begin{tabular}[htb]{lll}
\includegraphics[width=0.5\textwidth]{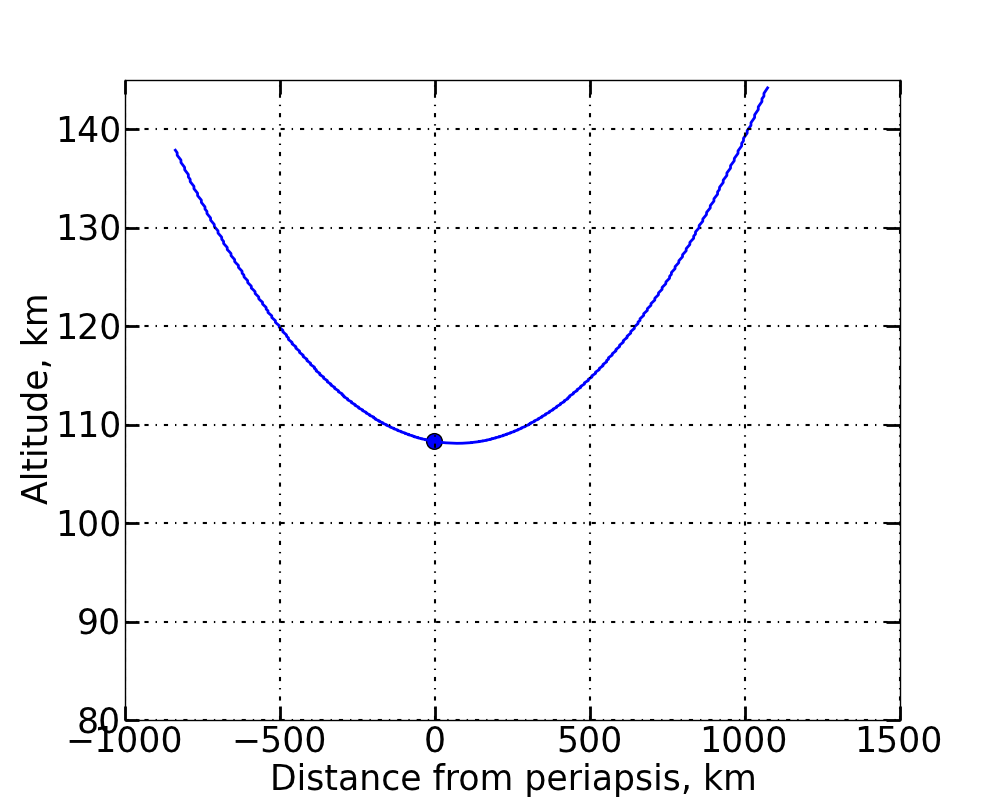}
\includegraphics[width=0.5\textwidth]{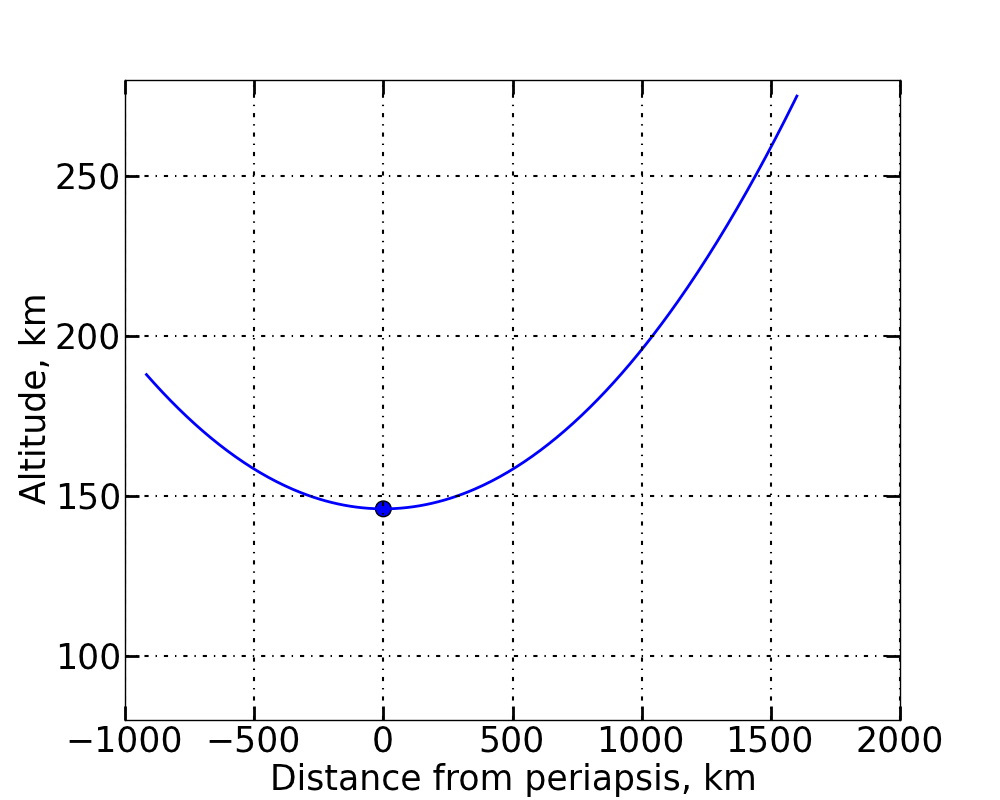} \\
\includegraphics[width=0.5\textwidth]{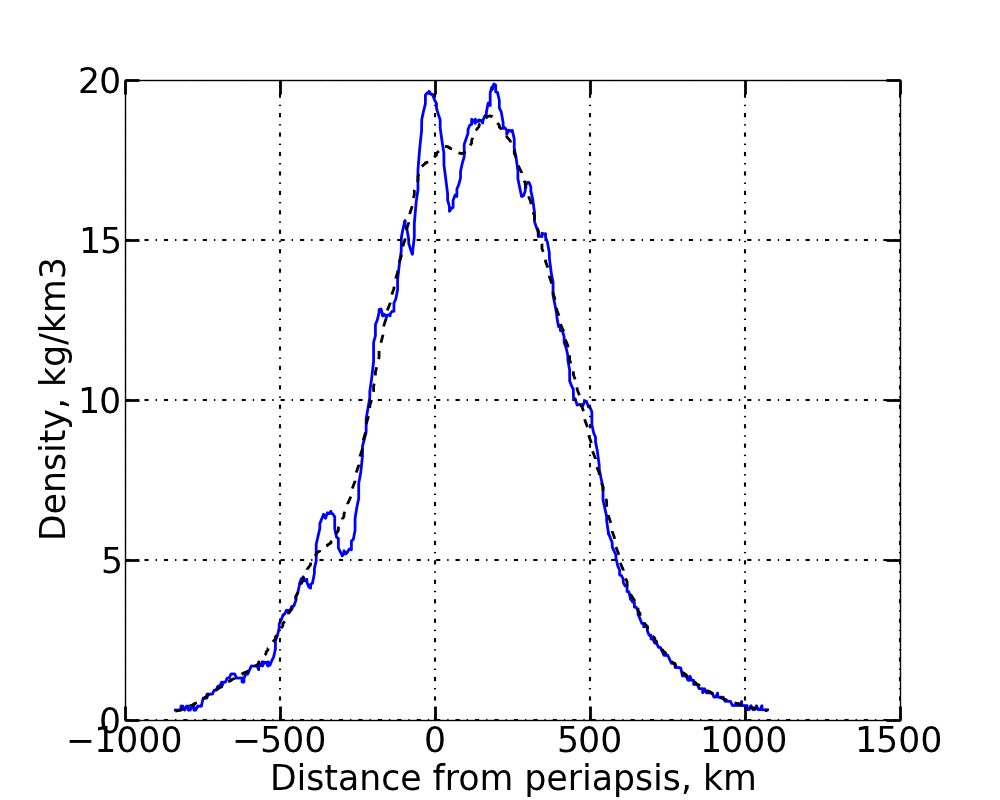}
\includegraphics[width=0.5\textwidth]{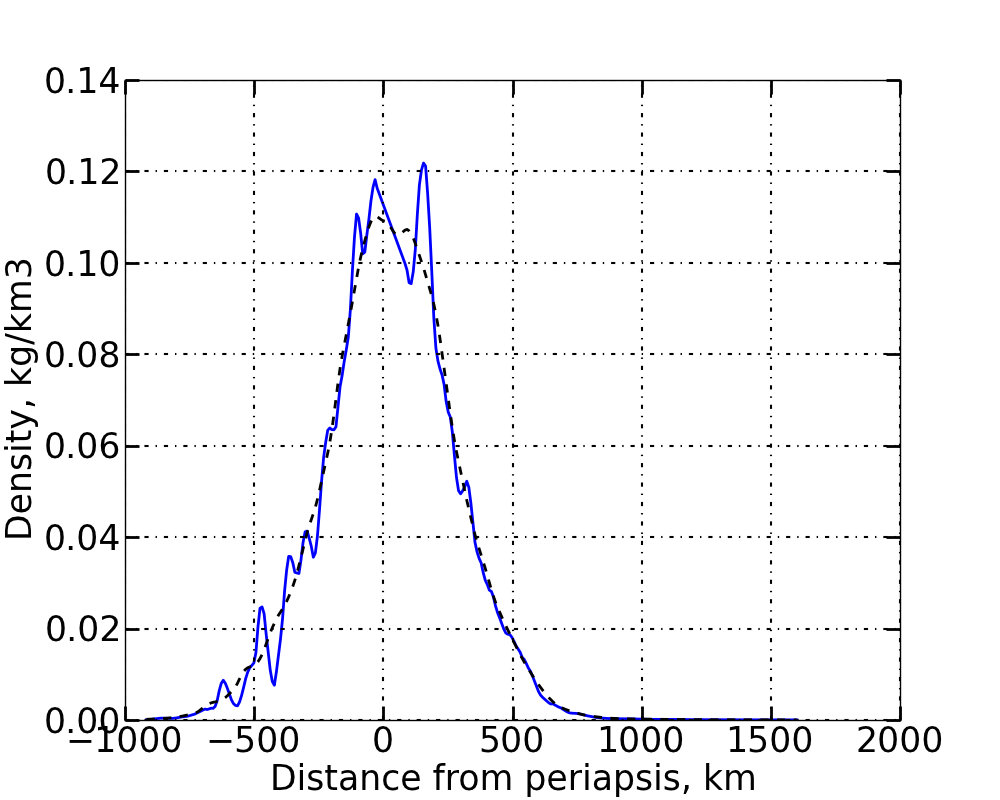} \\
\includegraphics[width=0.5\textwidth]{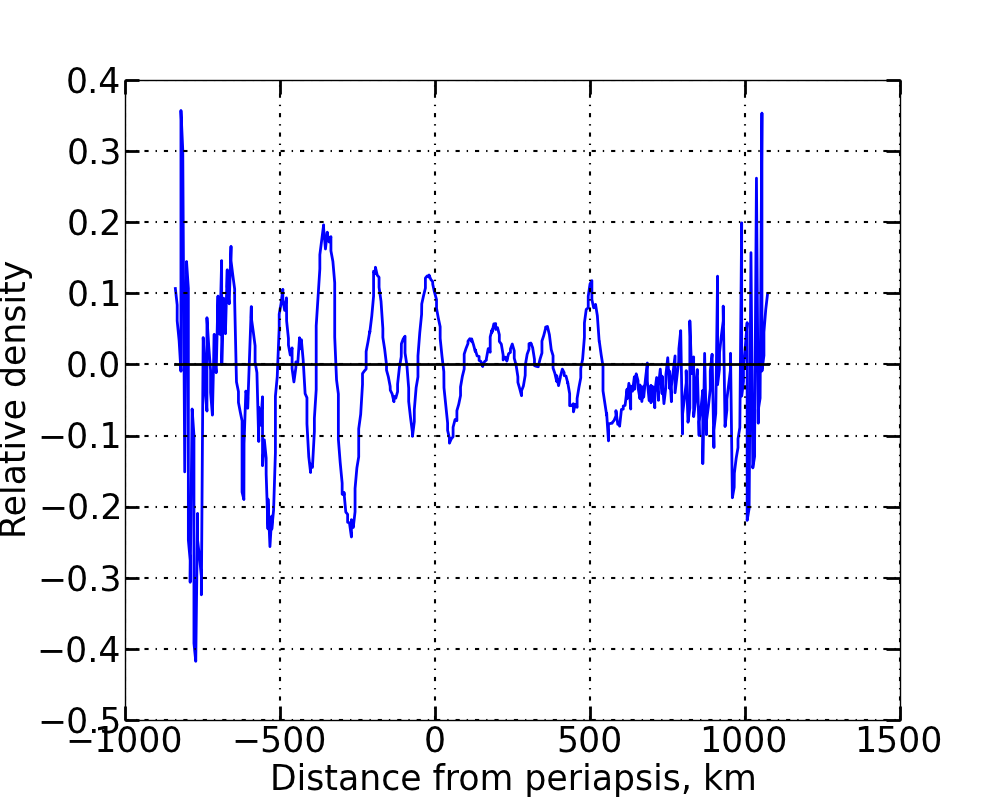}
\includegraphics[width=0.5\textwidth]{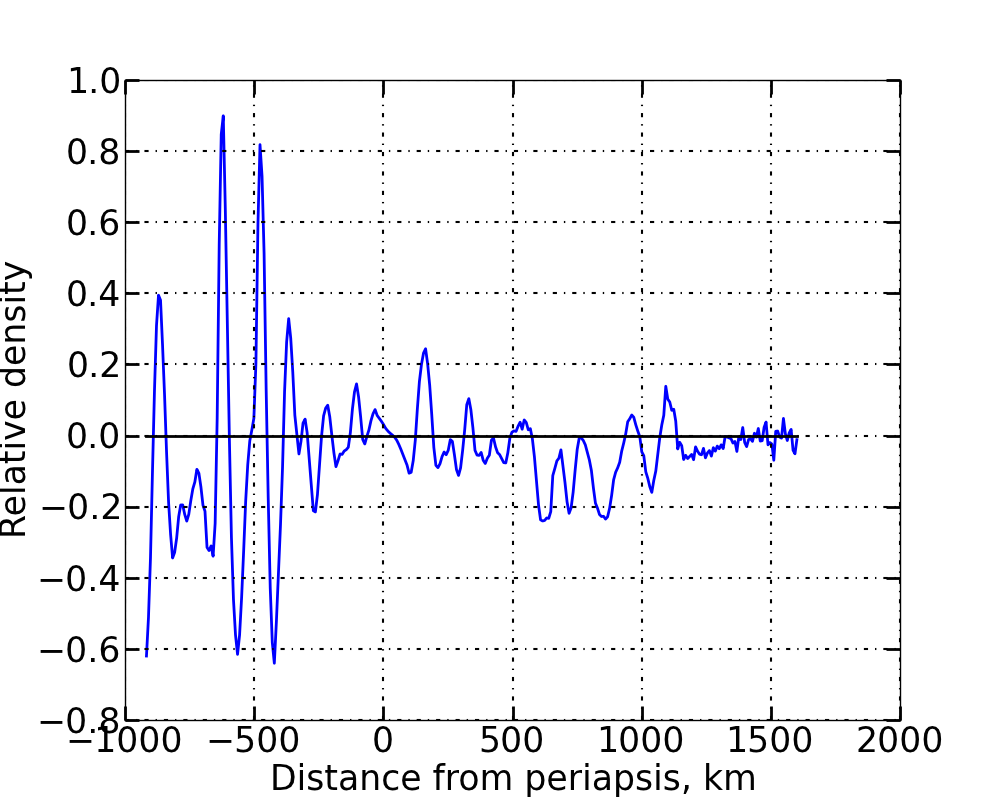} \\
\end{tabular}
\end{center}
\caption{Left: example of orbit 1046 from MGS, Right: example of orbit 3641 from MAVEN. From Left to Right: orbit's trajectory characterized by the displacement in altitude along the distance from periapsis in km; Density variations in kg~km$^{-3}$ in function of the distance from periapsis in km ; Relative density variation in function of the distance from periapsis in km.}
\label{ex_orbits}
\end{figure*}

Along each orbit trajectory, we extract the 
longitudes, 
latitudes, 
solar longitudes (L$_{s}$, which is the position of the planet on its orbit, defined as an angle from a reference position, corresponding by convention to the northern spring equinox), 
local times, 
altitudes, 
CO$_{2}$ density measurements, 
as well as 
the elapsed time from the periapsis. 
The geodesic distance from the periapsis 
is calculated from the latitude and longitude displacements. 
A relative density perturbation~$\delta \rho_{r}$ is obtained 
by subtracting 
the mean density $\rho_{m}$ 
\citep[considered here to be a 40-second rolling 
averaged density, as in][]{Tols:99,Tols:05,Tols:07,Tols:08,Crea:06ugw}
from
the instantaneous density $\rho_{i}$ ,
and by normalizing with the mean density
\begin{equation}
\delta \rho_{r} = \frac{\rho_{i} - \rho_{m}}{\rho_{m}} 
\label{reldens}
\end{equation}
\noindent Typical examples of 
orbit trajectory, 
absolute and relative density variations, 
obtained for the MGS orbit 1046 and for the MAVEN orbit 3641 
are shown in Figure \ref{ex_orbits}. 
Considering the relative density perturbations, 
rather than the absolute value,
enables a direct diagnostic of the effect of gravity waves, 
with the underlying assumption that 
the 40-second average provides an acceptable estimate of 
the ``background'' atmospheric state 
upon which the gravity waves propagate.

In order to quantify 
the amplitude (i.e. the intensity) 
of the observed gravity waves on a single orbit, 
and to assess the spatial and seasonal variability
of the gravity wave activity, 
we calculate for each orbit the Root Mean Square (RMS) of 
the fluctuations of relative densities~$\delta \rho_{r}$  
along the trajectory. 

\begin{figure*}[!ht]
 \begin{center}
\includegraphics[width=\textwidth]{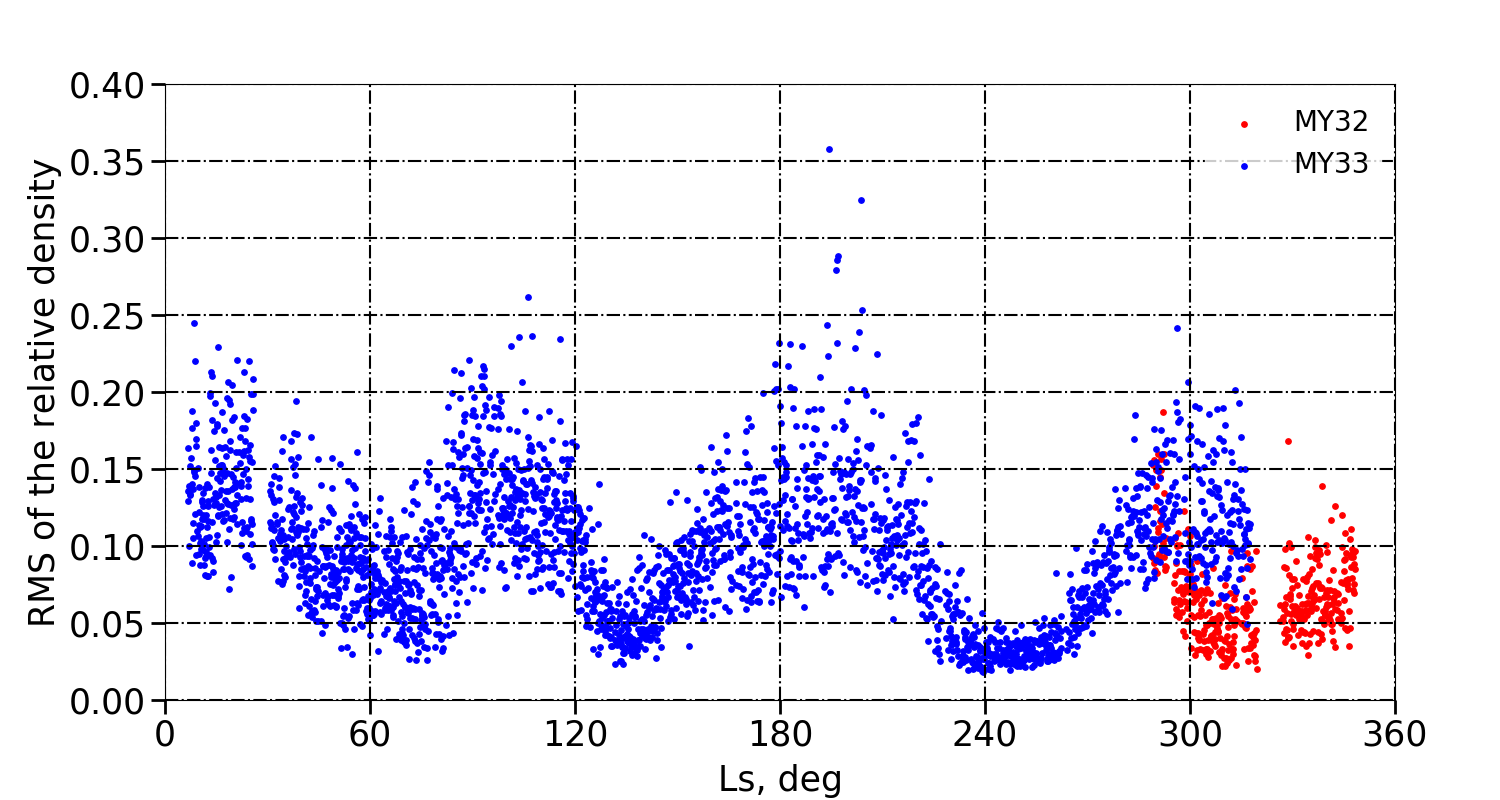}
\end{center}
\caption{Seasonal variability of GWs amplitudes measured by MAVEN/NGIMS. Each point corresponds to the RMS of the relative densities calculated over each orbit. In this figure the RMS has been calculated on the points around the periapsis, where the trajectory is close to be horizontal, at distances from the periapsis comprised between -700 and 700~km. This restriction reduces the altitude range to around 15~km above the periapsis. Data gathered from Martian Year (MY) 32 are in red, and data from MY33 in blue.}
\label{mvn_RMS_Ls}
\end{figure*}

\begin{figure*}[!ht]
 \begin{center}
\includegraphics[width=\textwidth]{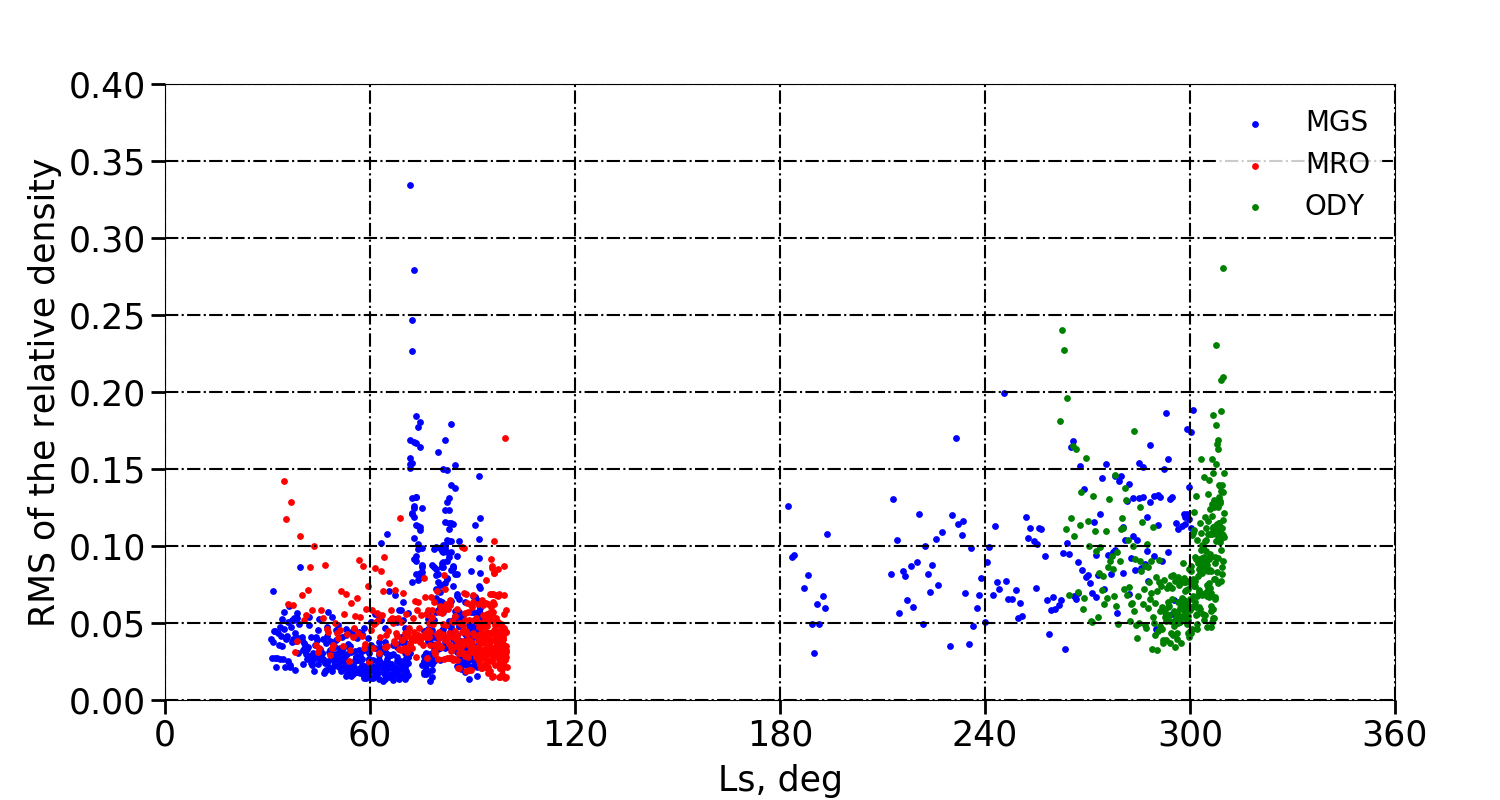}
\end{center}
\caption{Seasonal variability of GWs amplitudes measured by aerobraking instruments MGS, ODY and MRO. Each point corresponds to the RMS of the relative densities calculated over each orbit. In this figure the RMS has been calculated on the points around the periapsis, where the trajectory is close to be horizontal, at distances from the periapsis comprised between -400 and 400~km. Beyond these distances the aerobraking data become very noisy. This restriction reduces the altitude range to around 10~km above the periapsis.}
\label{aero_RMS_Ls}
\end{figure*}

Figure~\ref{mvn_RMS_Ls} (MAVEN/NGIMS data) and 
Figure~\ref{aero_RMS_Ls} (aerobraking data)
show the seasonal variations of the GW activity
as quantified by this RMS quantity, i.e. the RMS as a function of the L$_{s}$, all other parameters (longitudes, latitudes, local times, altitudes) confounded.
A distinctive pattern of amplitude fluctuations with season
is found in the MAVEN data in Figure~\ref{mvn_RMS_Ls}, 
in agreement with
the tendencies discussed in \citet{Tera:17}.


\subsection{Temperature estimates}

The background temperature $T$ is estimated at each point of each orbit with the ideal gas law and the hydrostatic equilibrium, as a function of the mean density of CO$_{2}$ $\rho$ and the altitude $z$ as follows:
\begin{equation}
{\int {\frac{\partial \rho}{\rho}}} = - \frac{g}{R_{CO_{2}} T}\int\partial z
\label{temperature}
\end{equation}
with $g$ the gravitational acceleration and $R_{CO_{2}}$ the ideal gas constant of CO$_{2}$.

We split the orbit in three parts.
\begin{enumerate}
\item The middle leg is the part of the orbit track close to the periapsis, where the displacement is almost horizontal, the density almost constant, and, consequently, where the temperature can no longer be deduced from equation \ref{temperature}. We arbitrarily define this middle leg as containing the points for which the ratio between the mean density and the maximal density is greater than 10\%.
\item The inbound leg corresponds to the points located "before" the perapsis not included in the middle leg.
\item The outbound leg refers to the points located "after" the periapsis not included in the middle leg.
\end{enumerate}
Thus the middle leg of the
measurements is excluded from 
the comparative analysis,
and we only keep the inbound
and outbound profiles for all aerobrakings and MAVEN/NGIMS measurements.

\begin{figure*}[!tb]
\begin{center}
\begin{tabular}[htb]{ll}
\includegraphics[width=0.5\textwidth]{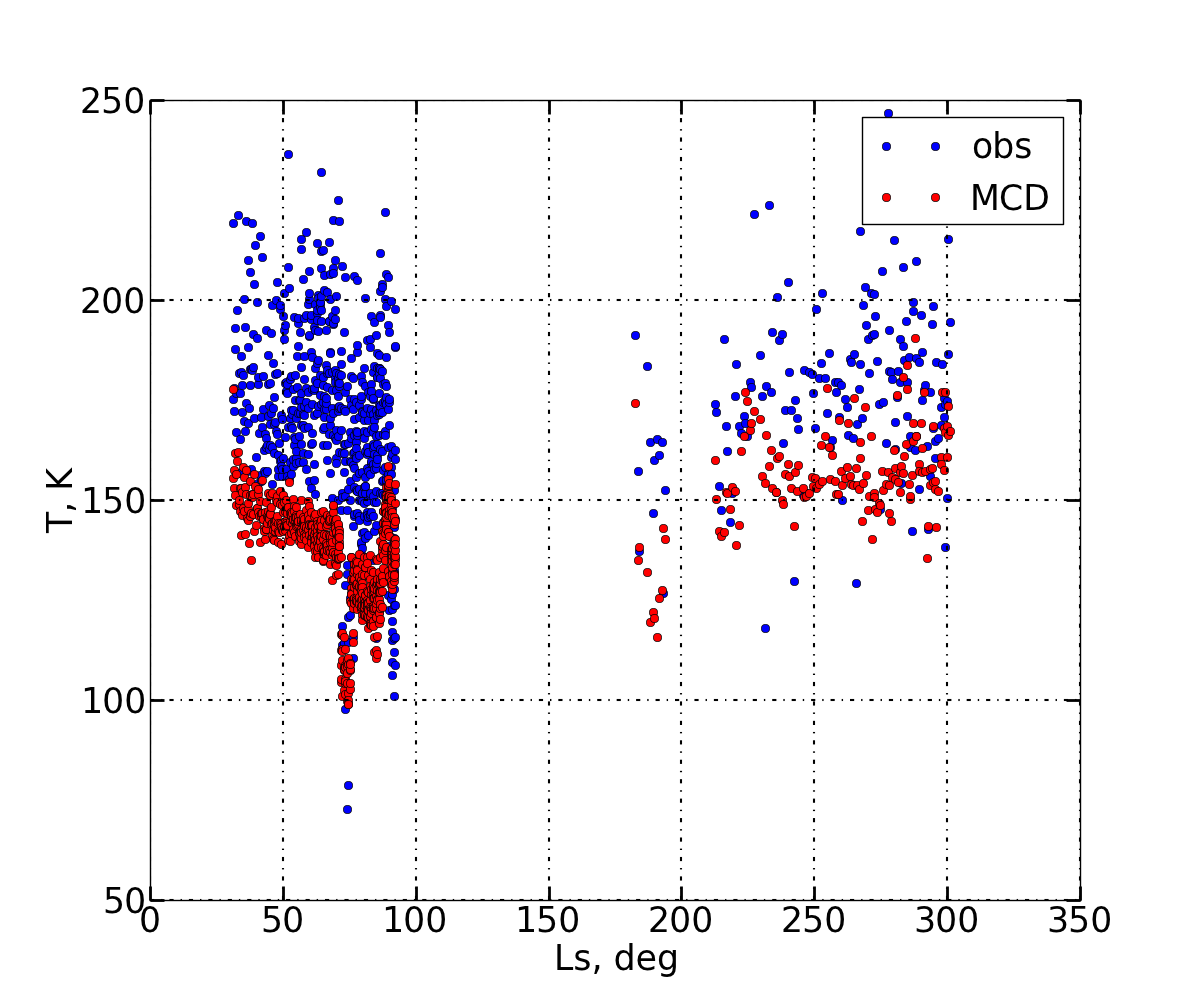} 
\includegraphics[width=0.5\textwidth]{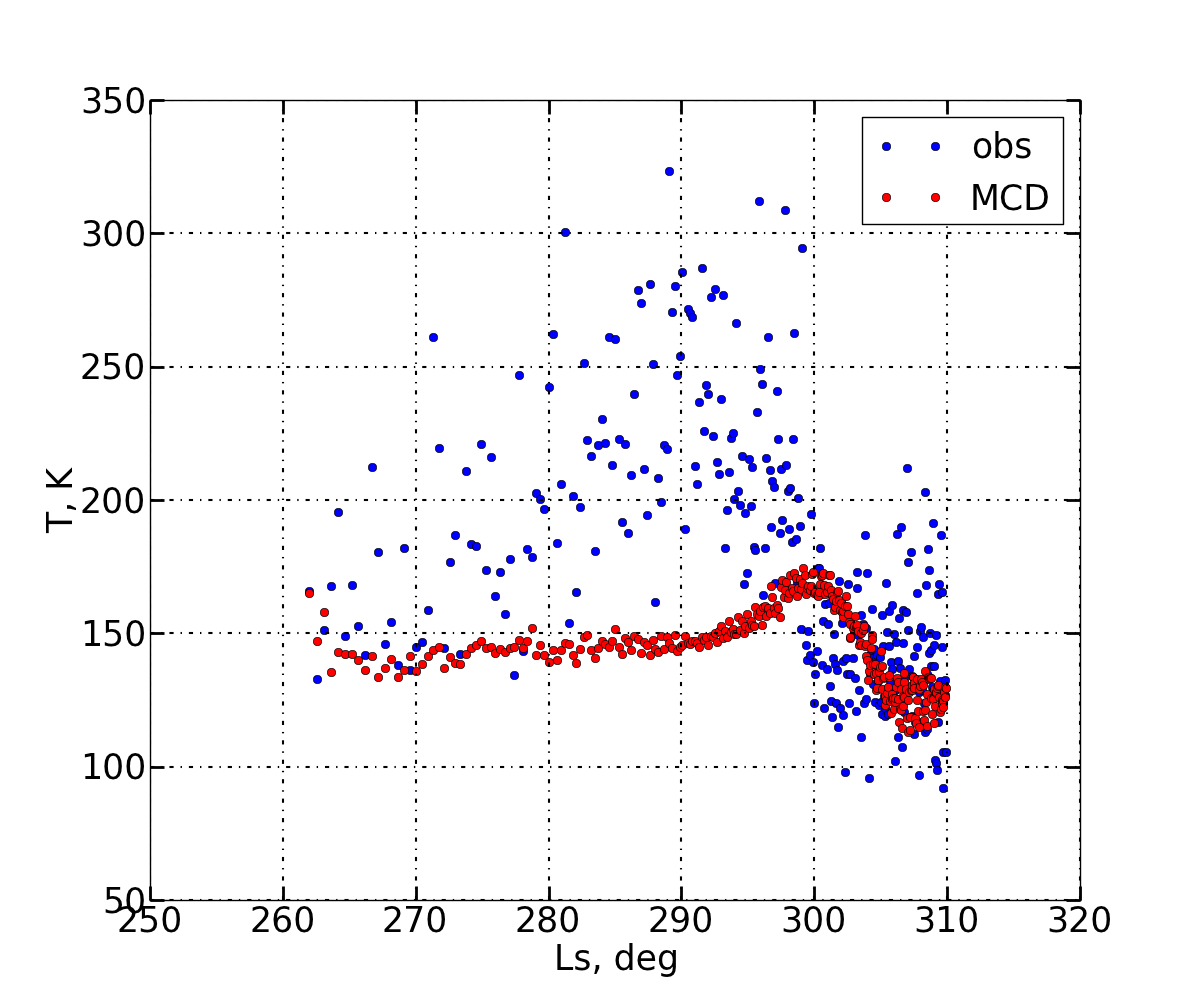}\\
{\centering\includegraphics[width=0.5\textwidth]{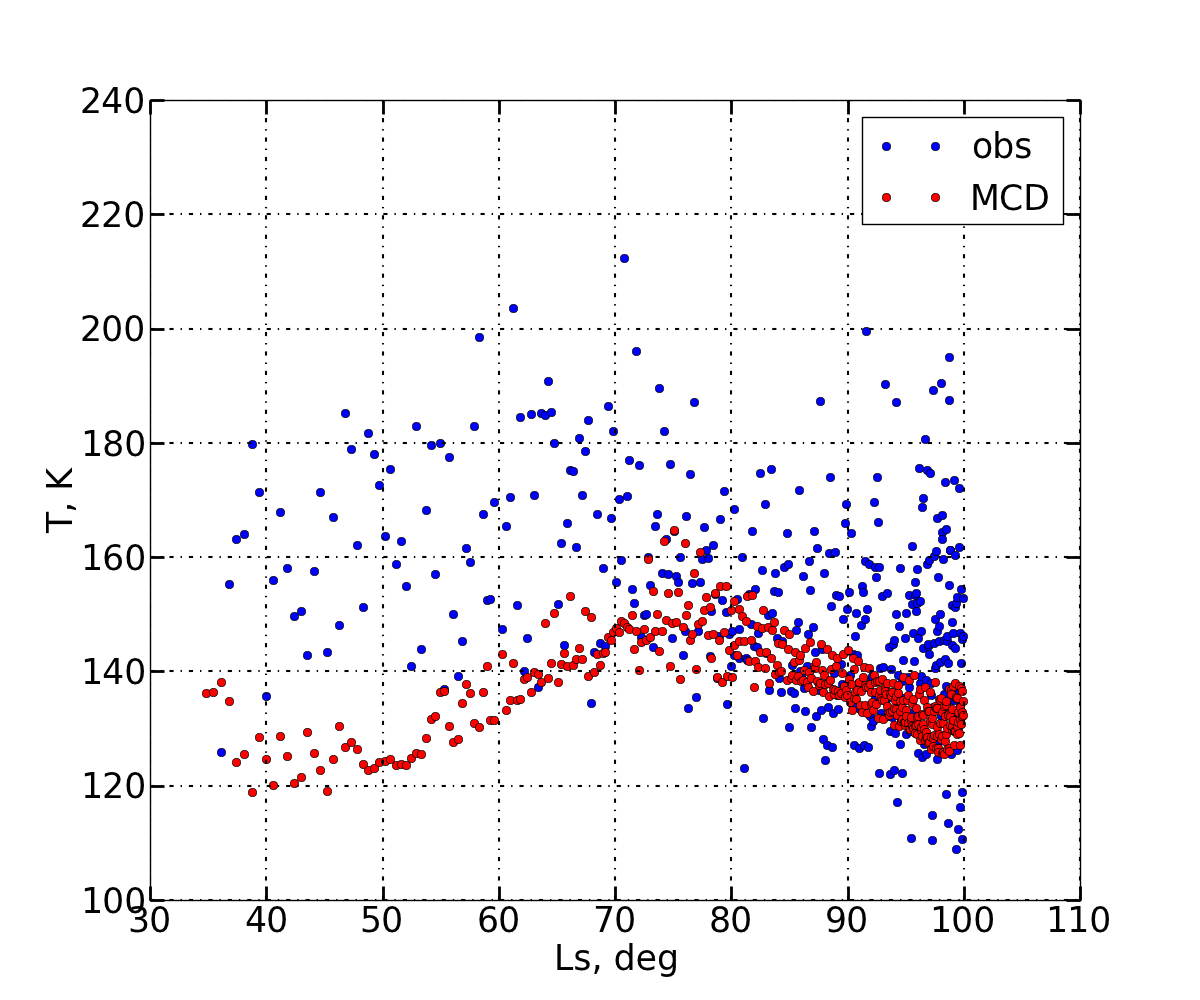}}
\end{tabular}
\end{center}
\caption{Mean background temperature estimated over the outbound leg and calculated from the CO$_{2}$ density observations (blue dots) and estimated with the MCD (red dots) as a function of Solar Longitude; from the upper to the lower : MGS, ODY and MRO}
\label{aerob_T}
\end{figure*}

\begin{figure*}[!tb]
\begin{center}
\includegraphics[width=\textwidth]{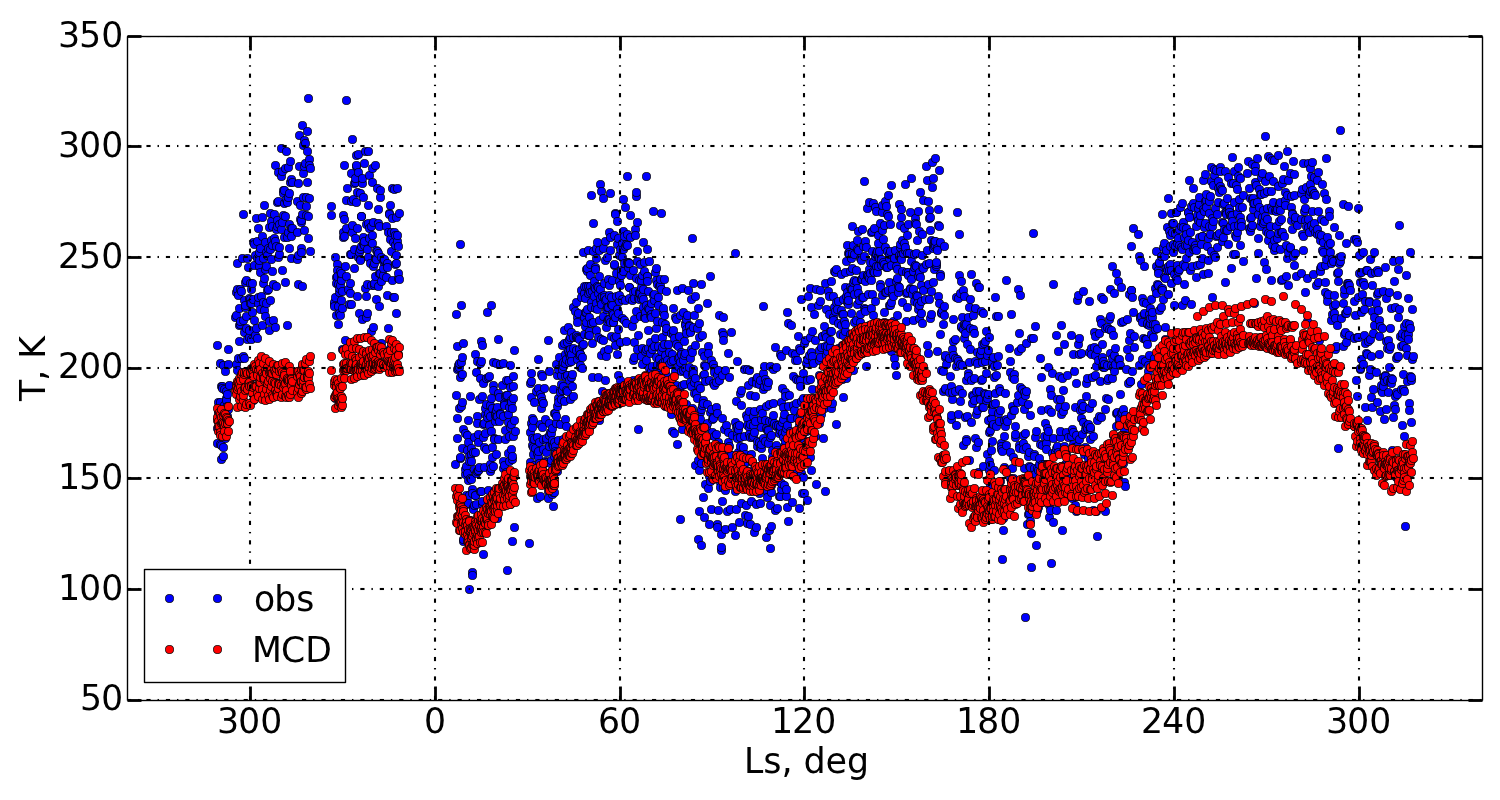}
\end{center}
\caption{Mean background temperature estimated over the outbound leg and calculated from the CO$_{2}$ density observations of NGIMS instrument (blue dots) and estimated with the MCD (red dots) as a function of Solar Longitude}
\label{mvn_T}
\end{figure*}

We found that in the inbound and outbound legs, 
the temperature profiles follow
a similar vertical gradient.
We thus study the variability
of temperature from one orbit to another
with a single representative value for both
the inbound and outbound legs,
chosen as the average value on each leg.
Those temperatures estimated 
from aerobraking and MAVEN/NGIMS measurements
are compared in Figure~\ref{aerob_T} and Figure~\ref{mvn_T}
with the temperature in the Mars Climate Database 
\citep[built from Global Climate Model (GCM) simulations][]{Mill:15}
for the same spatio-temporal coordinates
(L$_{s}$, 
longitude, 
latitude, 
altitude, 
local time).
Only the comparisons of 
temperatures measured on outbound legs 
versus 
temperature modeled in the MCD
are displayed for the sake of brevity;
the analysis for inbound legs is similar.
The MCD temperatures are systematically lower than
those observed by MAVEN and aerobraking,
and there is also much more variability 
in the observation data points; however, the overall
seasonal variability is well reproduced, except
the~$L_s = 290^{\circ}$ maximum observed by ODY.
This gives us confidence that
using a value of background temperature 
averaged over the inbound
and outbound legs is suitable
to carry out an analysis of 
the seasonal (climatological) trends.


\section{Vertical Propagation of Gravity Waves: analysis of the MAVEN observations in the thermosphere \label{sec:maven}}

\begin{figure*}[!ht]
 \begin{center}
\includegraphics[width=\textwidth]{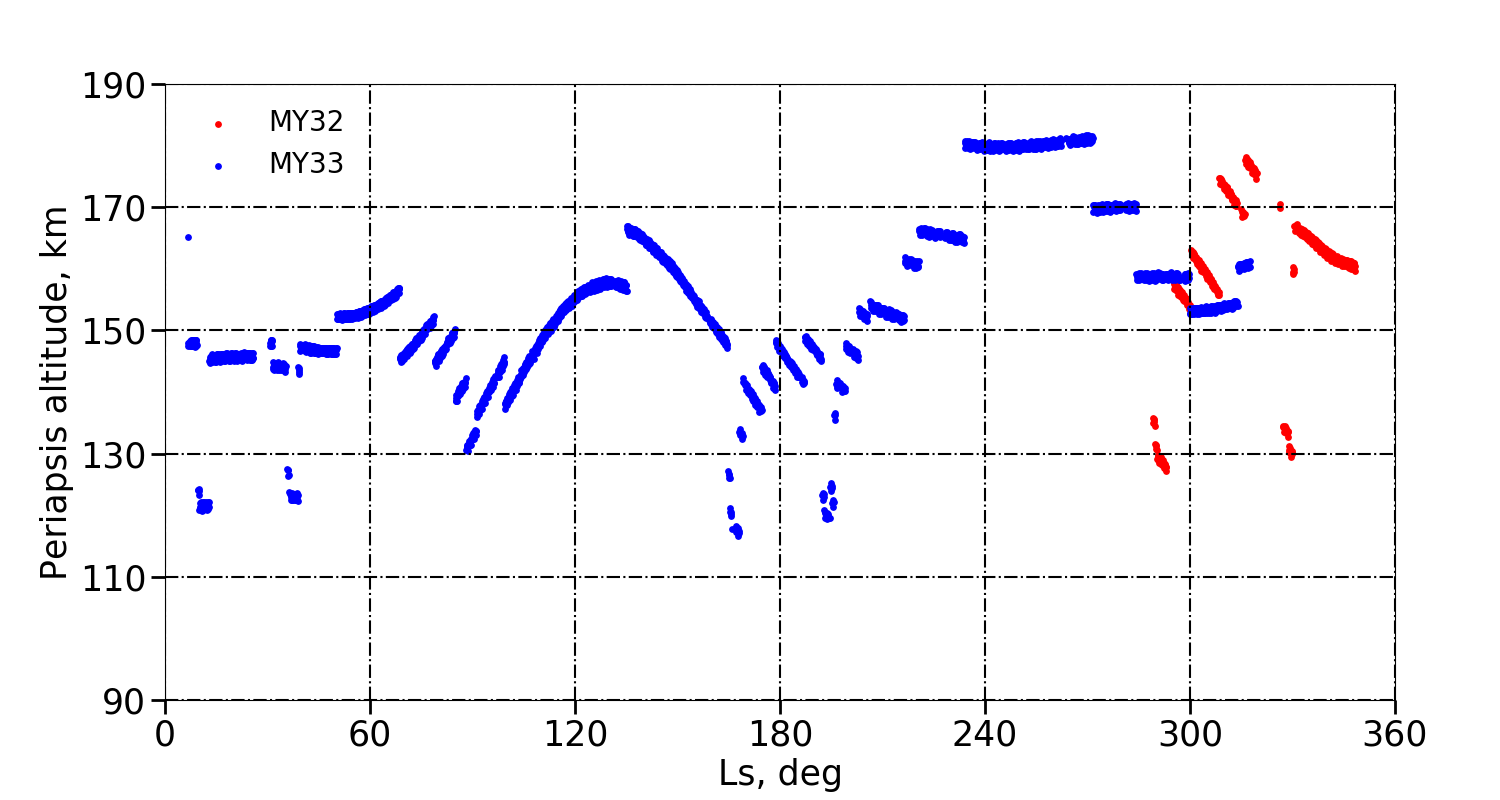} \\
\includegraphics[width=\textwidth]{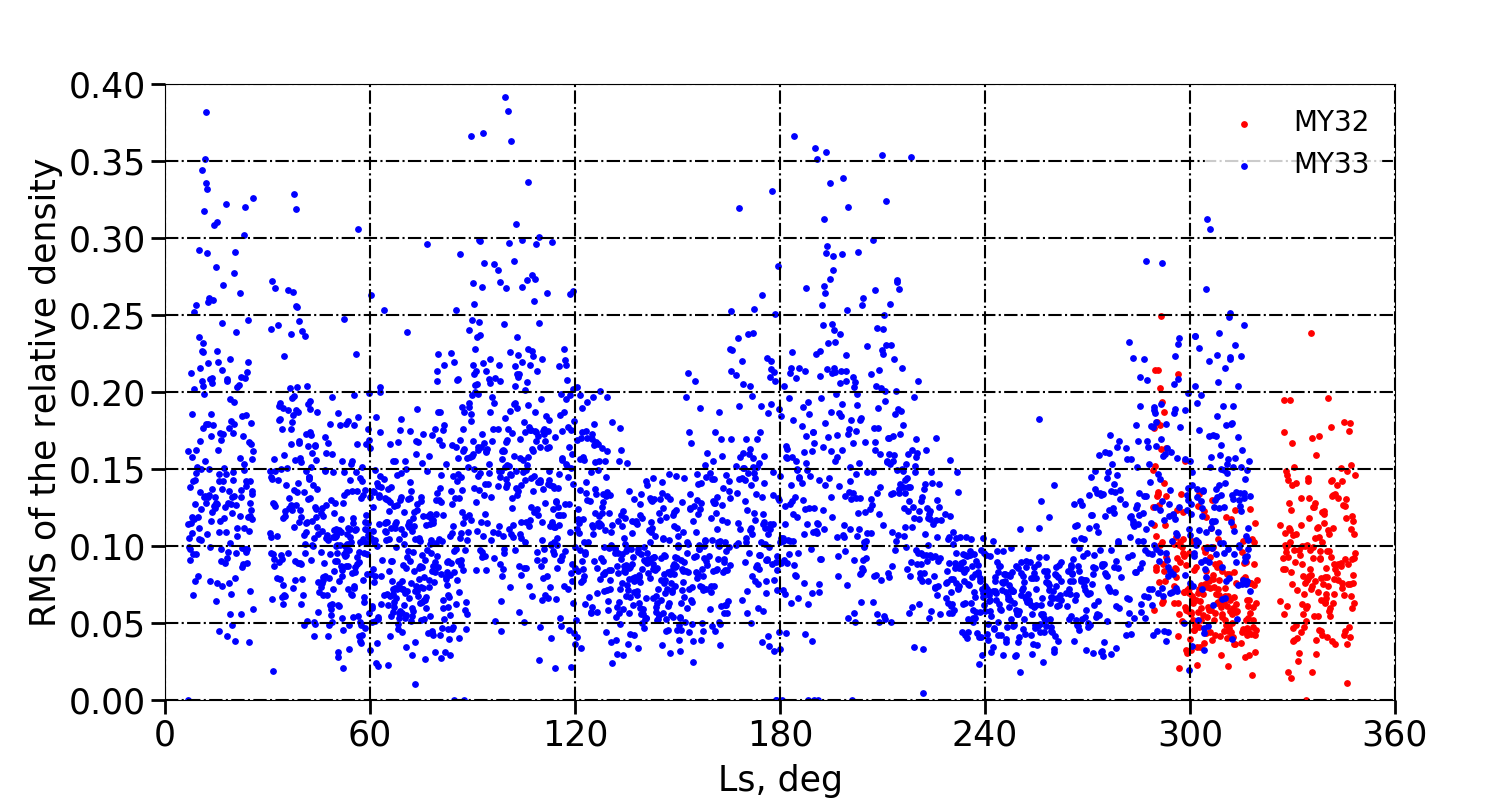}
\end{center}
\caption{Upper: Vertical (km) and seasonal (Solar Longitude in degrees) coverage of MAVEN (MVN) spacecraft, each dot corresponds to the periapsis location of one orbit. Down: Seasonal variability of GWs amplitudes measured by MAVEN/NGIMS at a constant pressure level P such as $4 \times 10^{-8} < P < 6 \times 10^{-8}$Pa. Each point corresponds to the RMS of the relative densities calculated over each orbit.  Data gathered from Martian Year (MY) 32 are in red, and data from MY33 in blue.}
\label{mvn_Pconst_rms_Ls_inbound}
\end{figure*}

In the absence of 
additional wave sources and
dissipation processes \citep[e.g., radiative damping][]{Ecke:11},
the amplitude of gravity waves is expected 
to grow exponentially with altitude 
as the atmospheric density decreases.
Conversely, the amplitudes of gravity waves 
appear to anti-correlate with altitude, 
according to 
the altitudes of the
MAVEN measurements shown in Figure~\ref{all_alt_Ls}
and the amplitudes~$\delta \rho$ of the perturbations
shown in Figure~\ref{mvn_RMS_Ls}.
In other words, in the MAVEN observations,
gravity-wave amplitude seems
to correlate with density,
as opposed to an anti-correlation
expected if the amplification of gravity-wave
amplitude with altitude (and reduced density)
was the only controlling factor.
This is confirmed by considering
the seasonal variations of density perturbations~$\delta \rho$
at a constant pressure level, 
e.g. at pressures~$4 \times 10^{-8} < P < 6 \times 10^{-8}$~Pa (corresponding to altitudes between $\sim$160 and $\sim$240~km)
in Figure \ref{mvn_Pconst_rms_Ls_inbound}. 
The observed variability in gravity-wave amplitude
must be controlled
by either the sources of those waves
and/or the impact of saturation and critical levels.

In the MAVEN measurements,
gravity wave activity in the thermosphere
is randomly distributed with longitude and latitude
(figures not shown).
No correlation appears to exist 
between this gravity wave activity
and either the position of 
topographical highs and lows (mountains and craters),
or the position of mesospheric jet-streams.
This suggests that 
the regional distribution of the intensity of gravity waves
is more controlled 
by propagation effects 
\citep[e.g., filtering by saturation or critical levels,][]{Frit:03}
than 
by the distribution of the sources triggering those waves.

The background horizontal wind plays a particularly crucial role
in impacting the conditions 
for the upward propagation of gravity waves
emitted in the troposphere.
A critical level occurs
when and where 
the background horizontal wind velocity~${\bar{u}}$ 
almost equals the gravity wave phase speed~$c$ 
\citep[first Eliassen-Palm theorem,][]{Lind:81}. 
A gravity wave that reaches a critical level 
can no longer propagate towards the thermosphere:
hence horizontal circulations may filter out
gravity waves emitted in the troposphere from
the mesosphere and the thermosphere.
%
%

Considering, for the sake of simplicity, 
a gravity-wave phase speed~$c=0$
(typical of orographic gravity waves),
we explored 
the regional and seasonal variability of 
background horizontal winds~${\bar{u}}$ 
simulated in the MCD from
the troposphere to the lower mesosphere
(since no measurements of such winds
are available).
We found no correlation between this variability,
and the regional and seasonal variability
of the gravity wave amplitudes observed by MAVEN (not shown).
While the modeled winds have not been validated and may differ from reality, there is no reason to explain
the variability of the observed gravity wave amplitudes
solely with the occurrence of critical levels.

It follows from the above discussions
that the most likely
possibility to explain
the observed variability of gravity wave amplitude 
in the MAVEN observations
is the breaking/saturation due to convective instability.
This shall lead to, according to \citet{Tera:17},
the gravity wave amplitudes to be 
inversely proportional to the background temperature.
Let us propose an alternate, yet equivalent,
derivation of the theoretical arguments in \citet{Tera:17}
that we will use in section~\ref{sec:aero}.

The saturation of a gravity wave occurs 
as soon as it encounters 
convective instability \citep{Lind:81,Hauc:87,Tera:17}.
Local mixing occurs as the gravity wave breaks,
inducing an adiabatic (neutral) temperature lapse rate.
We consider the case of a 
medium-frequency gravity wave
${f} \ll {\omega} \ll {N}$, 
where~${f}$, ${\omega}$ and ${N}$ 
are respectively 
the Coriolis, 
the gravity-wave and 
the Brunt-V\"{a}is\"{a}l\"{a} 
frequencies, with~$N$ such that
\[
N^{2} = \frac{g}{T} \left[ \frac{\partial T}{\partial z} + \frac{g}{C_{p}} \right]
\]
assuming the short-wavelength approximation
${2\,H\,k_{z}} \gg {1}$,
where ${k_{z}}$ is the vertical wave number.
Which are reasonable assumptions 
for most gravity waves 
observed in planetary upper atmospheres
\citep{Frit:03}.
In those conditions,
according to~\citet{Hauc:87},
the saturated conditions lead to
\begin{equation}
{k_{z}}{\theta'_{s}} = \frac{N^{2} \bar{\theta}}{g}
\qquad
\Rightarrow
\qquad
{\frac{\theta'_{s}}{\theta}} = \frac{N^{2}}{g k_{z}}
\label{kz}
\end{equation}
\noindent where ${\theta'_{s}}$ is the amplitude of the wave at saturation
(expressed in perturbations of potential temperature), $\bar{\theta}$ 
the background potential temperature and~$g$ the acceleration of gravity. 
Besides, the linearized fluid equations applied to 
the propagation of gravity waves \citep{Frit:03}
lead to:
\begin{equation}
\frac{\theta'}{\bar{\theta}} = \frac{1}{c_s^2} \frac{P'}{\bar{\rho}} - \frac{\rho'}{\bar{\rho}}
\label{cs1}
\end{equation}
\noindent where $\rho$ is the density, $P'$ and $\rho'$ the pressure and density perturbations, and ${c_{s}}$ the sound speed. We can neglect the compressibility term related to the background density gradient, which is equivalent to filter out acoustic gravity waves (${c_{s}} \rightarrow \infty$). This entails:
\begin{equation}
{\left| \frac{\rho'}{\bar{\rho}} \right|} = {\left| \frac{\theta'}{\bar{\theta}} \right|}
\label{cs2}
\end{equation}
Combining equations~\ref{kz} and~\ref{cs2}, we obtain the 
equation expressing the relative density perturbations by gravity waves:
\begin{equation}
\boxed{
\delta \rho = \frac{|\rho'|}{\bar{\rho}} = \frac{N^{2}}{k_{z} g}
}
\label{saturation2}
\end{equation}
\noindent which corresponds to the observed diagnostic described in equation~\ref{reldens}.
Isothermal background profiles~$T=T_0$ are often observed in the
Martian thermosphere, where EUV heating is offset by molecular conduction \citep{Boug:90}.
In the specific case of isothermal profiles, ${N^{2}}$ can be reduced to:
\begin{equation}
N^{2} 
= \frac{g}{\bar{\theta}} \frac{\textrm{d}\bar{\theta}}{\textrm{d}z} 
= \frac{g^{2}}{C_{p} T_{0}}
\label{N}
\end{equation}
which yields the ``inverse temperature'' dependency 
\citep{Tera:17}
in the case of isothermal profiles at saturation:
\begin{equation}
\boxed{
\delta \rho =
\frac{|\rho'|}{\bar{\rho}} = \frac{g}{k_{z} \, C_{p}} \frac{1}{T_{0} } 
}
\label{saturation3}
\end{equation}

\begin{figure*}[!ht]
 \begin{center}
\includegraphics[width=\textwidth]{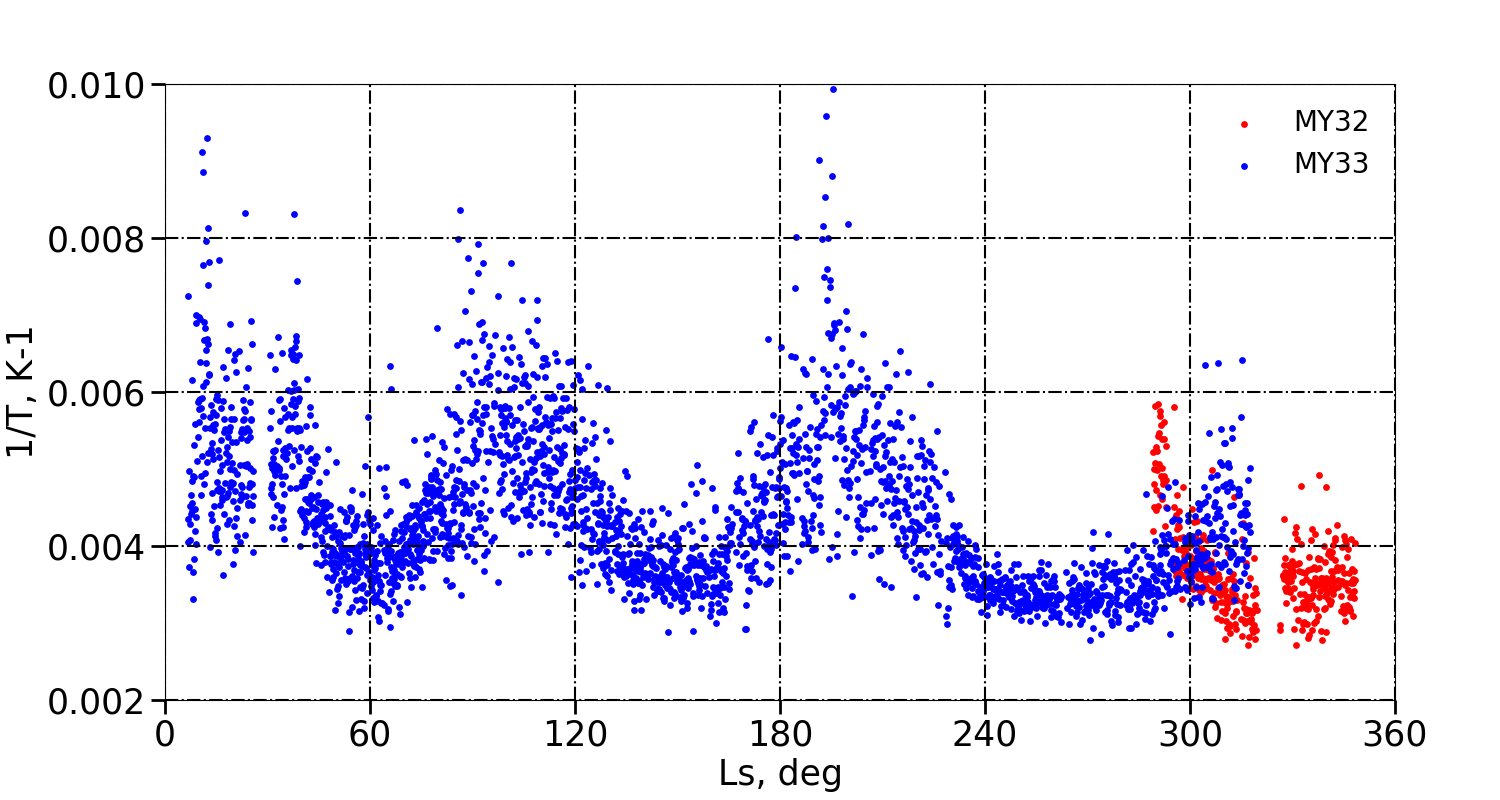}
\end{center}
\caption{Seasonal variability of the background temperature estimated from MAVEN/NGIMS density measurements (ideal gas law and hydrostatic equilibrium). Each point corresponds to the inverse of the mean background temperature calculated over the outbound leg of each orbit. Data gathered from Martian Year (MY) 32 are in red, and data from MY33 in blue.}
\label{mvn_T-1_Ls}
\end{figure*}

\begin{figure*}[!ht]
 \begin{center}
\includegraphics[width=\textwidth]{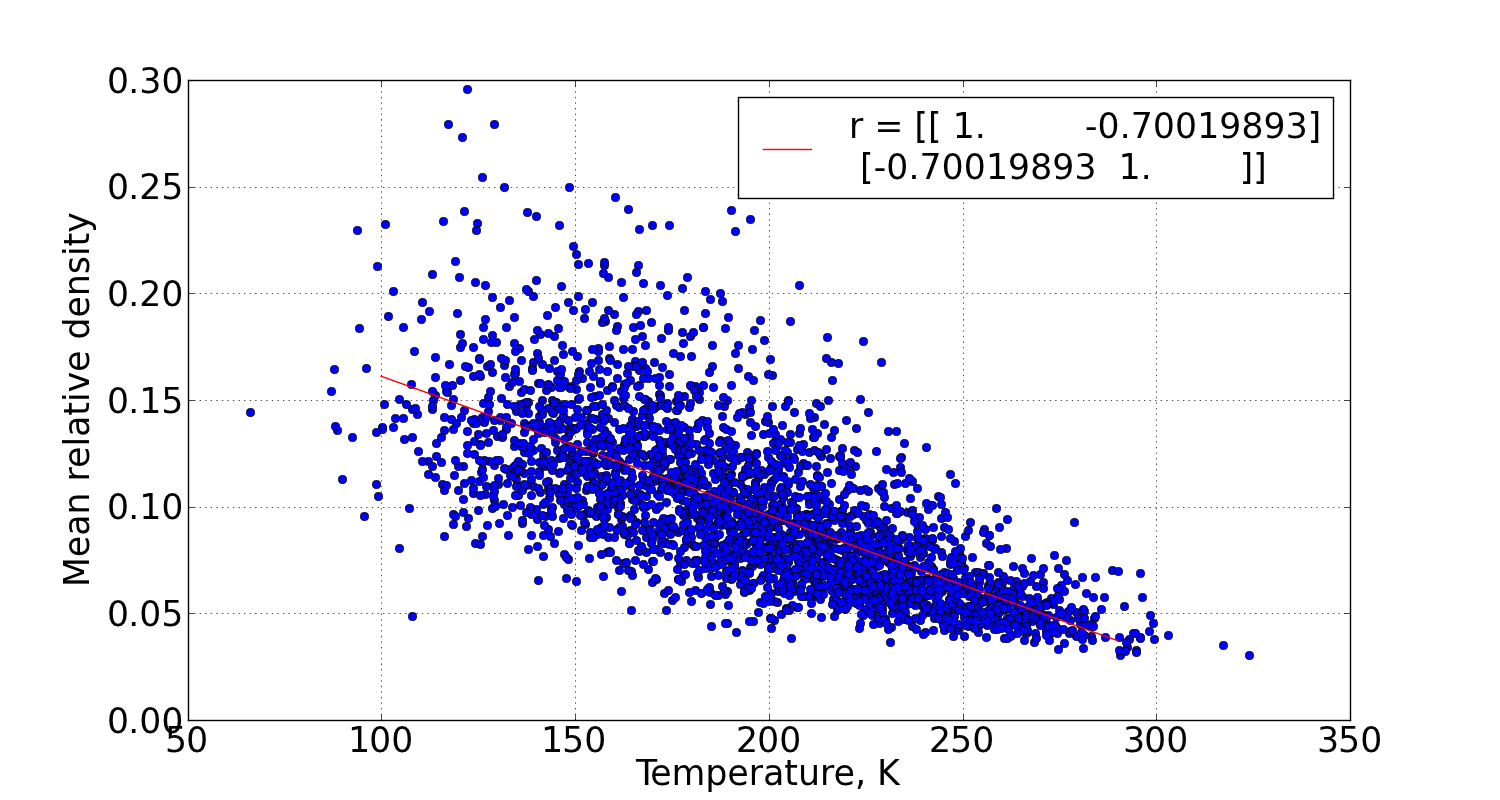}
\end{center}
\caption{Correlation between the average of the absolute relative density and the average of the background temperature calculated for MAVEN/NGIMS data over the inbound leg of each orbit. Temperature is obtained from the density observations by means of the ideal gas law and the hydrostatic equilibrium}
\label{reldens_T_correlation}
\end{figure*}

MAVEN data are acquired high 
in the Martian thermosphere (above 150~km) 
even for deep dip acquisitions: 
hence the temperature profiles retrieved by MAVEN
are approximately isothermal \citep{Engl:17,Tera:17}.
The temperature profiles modeled and compiled in the MCD
also indicate widespread isothermal profiles
at the altitudes probed by MAVEN.
Comparing Figures \ref{mvn_RMS_Ls} and \ref{mvn_T-1_Ls}
confirms qualitatively 
equation~\ref{saturation3}, i.e. the 
correlation between 
the amplitude of gravity wave perturbations
and 
the inverse background temperature.
Quantitatively, 
in the case of the inbound leg of each orbit, 
a correlation coefficient $R \simeq 0.70$ 
between the average of the relative density and 
the calculated temperature is found 
(see Figure \ref{reldens_T_correlation}). 
Our analysis of the MAVEN is thus compliant 
with the one conducted by \citet{Tera:17},
and we now turn to the analysis of 
aerobraking data in the lower thermosphere.

\section{Gravity Waves in the Lower Thermosphere: Aerobraking Data \label{sec:aero}}

\subsection{Analysis}


Aerobraking data have been studied in the past to observe the activity of gravity waves in the lower thermosphere, either to discuss the variability of potential sources \citep{Crea:06ugw} or to assess wave filtering by zonal jets and how large-amplitude GWs could penetrate to high altitudes  
\citep{Frit:06}. Here we assess if the ``inverse temperature'' correlation inferred from the MAVEN/NGIMS data \citep[][and section~\ref{sec:maven} of this paper]{Tera:17} can be extended to those lower-thermosphere aerobraking observations obtained by the three accelerometers of MGS, ODY and MRO.

\begin{figure}[!tb]
\begin{center}
\begin{tabular}[htb]{ll}
\includegraphics[width=6.cm,clip=t]{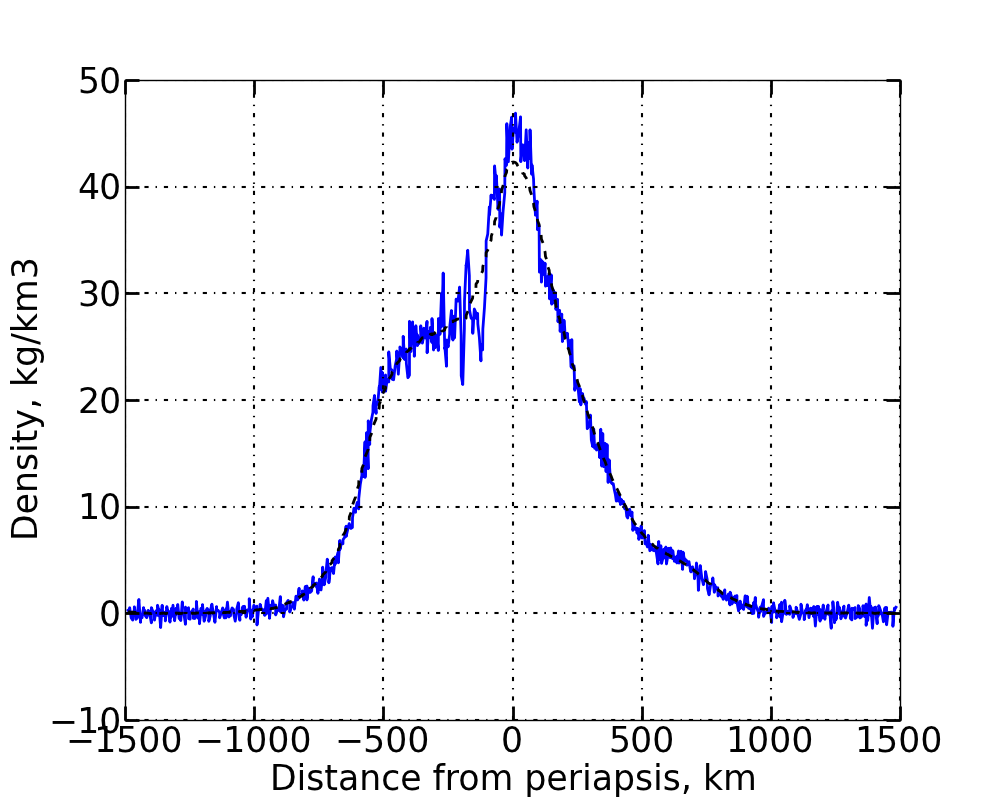}
&
\includegraphics[width=6.cm,clip=t]{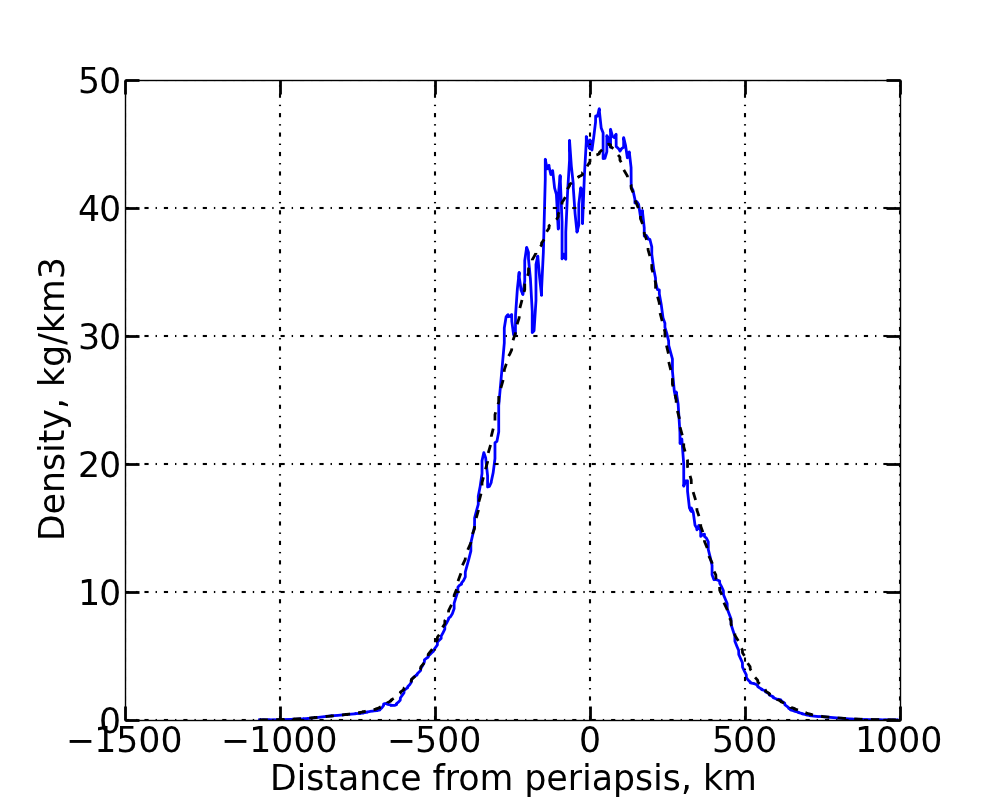}\\
\end{tabular}
\end{center}
\caption{Examples of orbit 155 of ODY and orbit 250 of MRO. Density variations in kg~km$^{-3}$ in function of the distance from periapsis in km }
\label{polar_vortex}
\end{figure}

In the aerobraking observations, as is emphasized by \citet{Tols:05} and \citet{Tols:08}, the intensity of density perturbations are systematically lower when the spacecraft enters the polar vortex (e.g. MRO during the southern hemisphere winter and ODY during the northern hemisphere winter). Figure \ref{polar_vortex} shows two examples: 
ODY orbit 155, which goes through the northern hemisphere winter vortex at $Ls=298.30^{\circ}$ and latitude 82.43$^{\circ}$N, 
and MRO orbit 250, going through the southern hemisphere winter vortex at $Ls=90.01^{\circ}$ and latitude=69.50$^{\circ}$S. 
These variations of density perturbations within the same orbital track 
could be
explained by the anti-correlation between temperature and gravity wave activity explained above \citep[an explanation that was not provided in][]{Tols:08}. Polar warming at thermospheric altitudes \citep[first observed by ODY during aerobraking,][]{Keat:03} results from the adiabatic heating generated by the subsidence of air over the winter pole produced by strong interhemispheric transport \citep{Gonz:09b}. The entry of the spacecraft inside the polar vortex is then expected to be associated with an increase of temperature, leading to a decrease of gravity wave activity according to equation~\ref{saturation3}.

\begin{figure*}[!tb]
\begin{center}
\begin{tabular}[htb]{ll}
\includegraphics[width=0.5\textwidth]{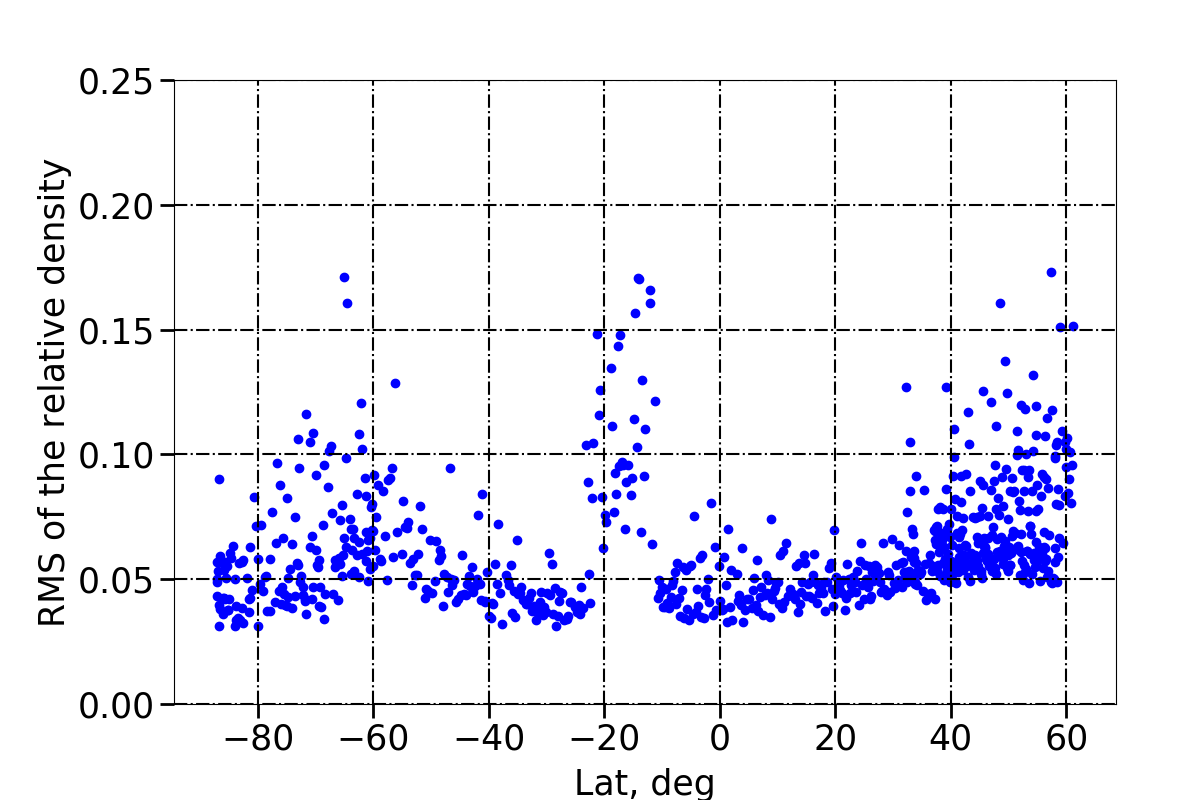}
\includegraphics[width=0.5\textwidth]{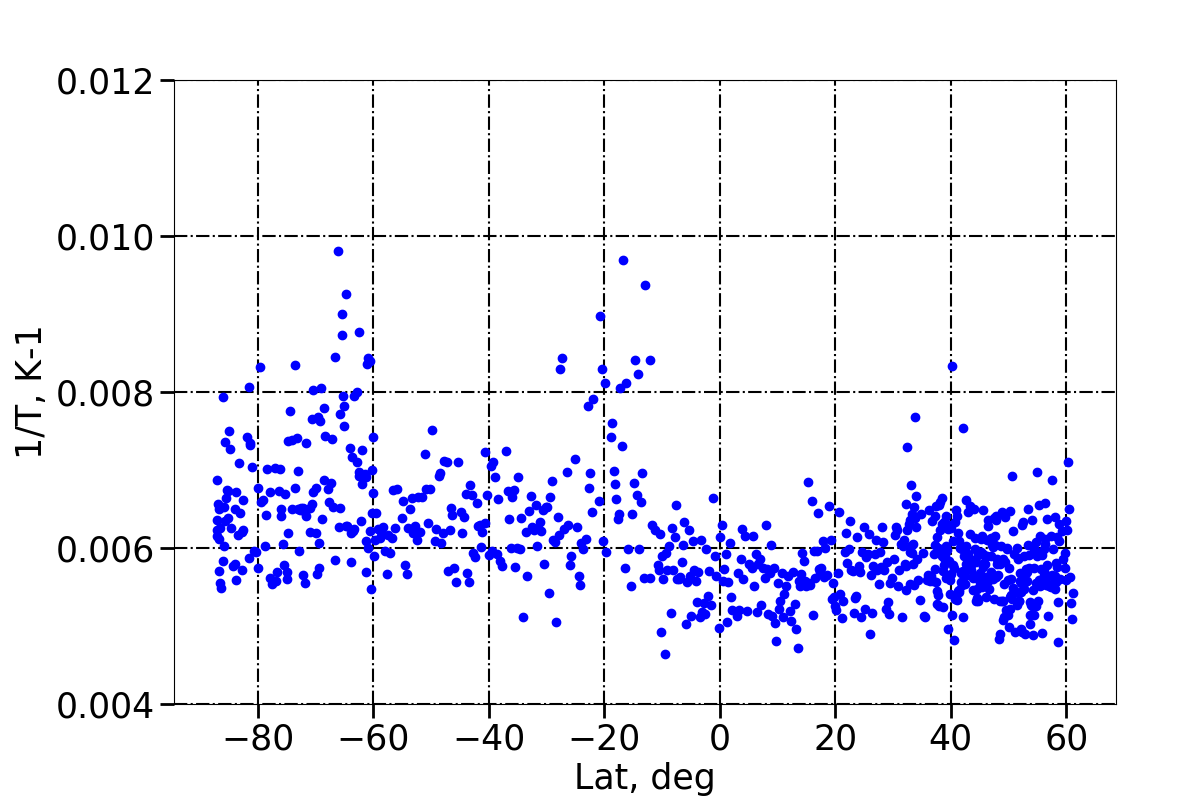}\\
\includegraphics[width=0.5\textwidth]{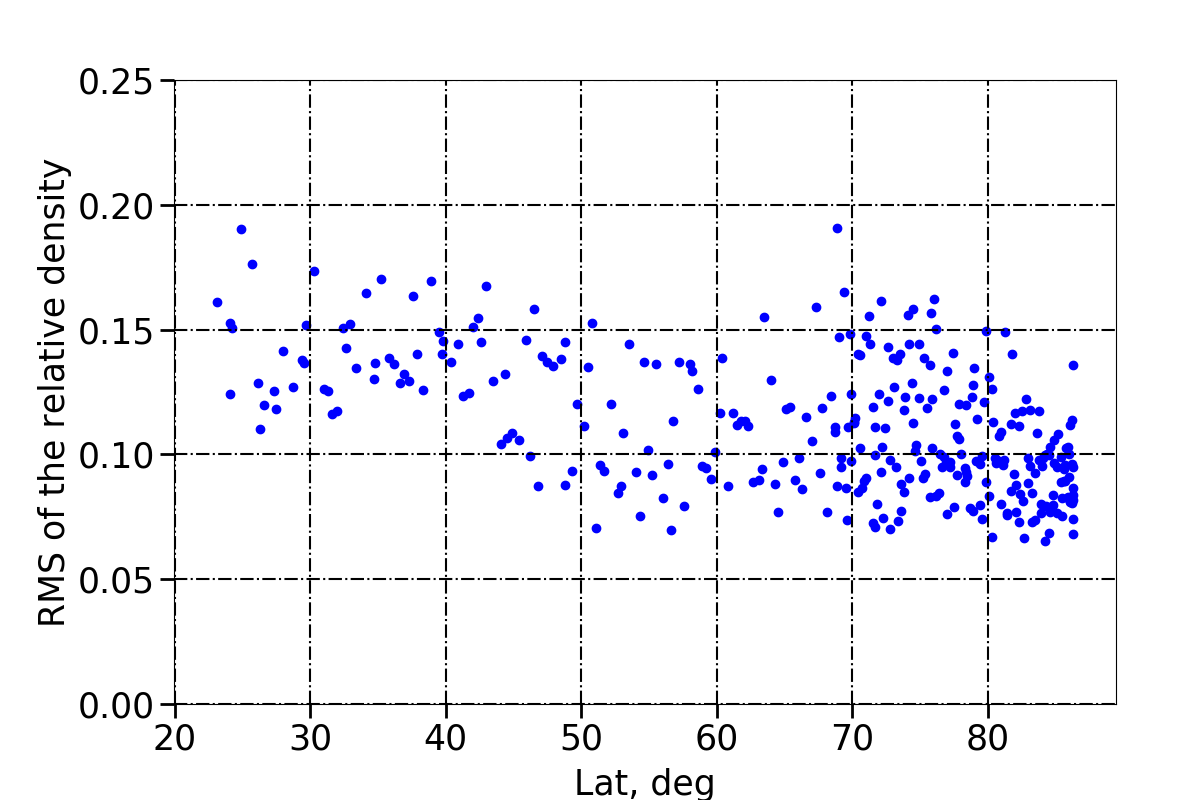}
\includegraphics[width=0.5\textwidth]{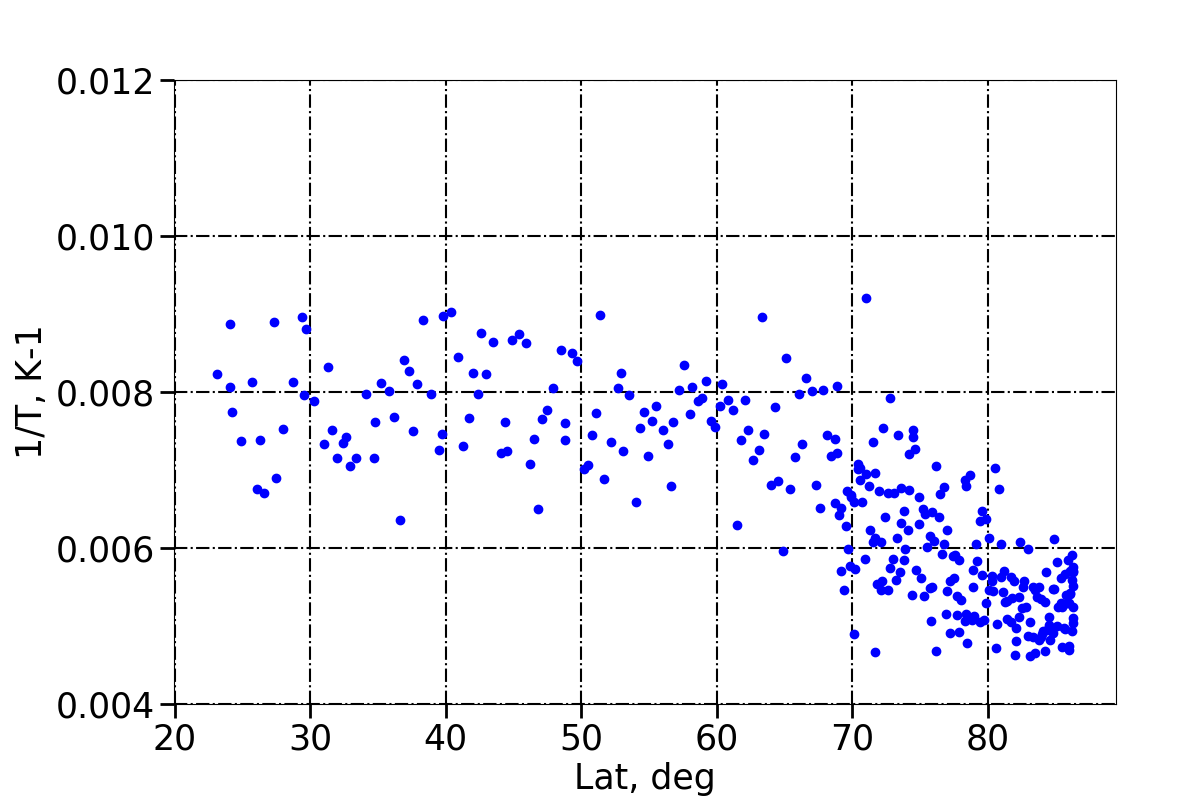}\\
\includegraphics[width=0.5\textwidth]{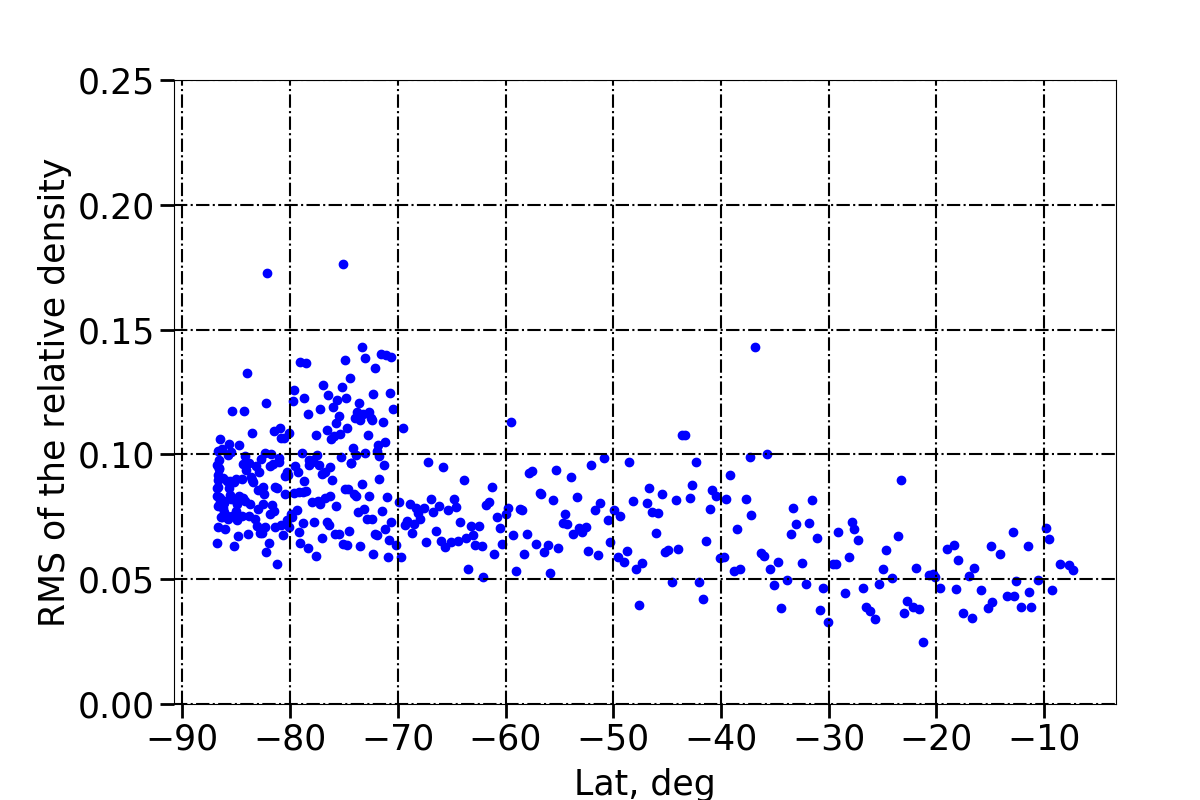}
\includegraphics[width=0.5\textwidth]{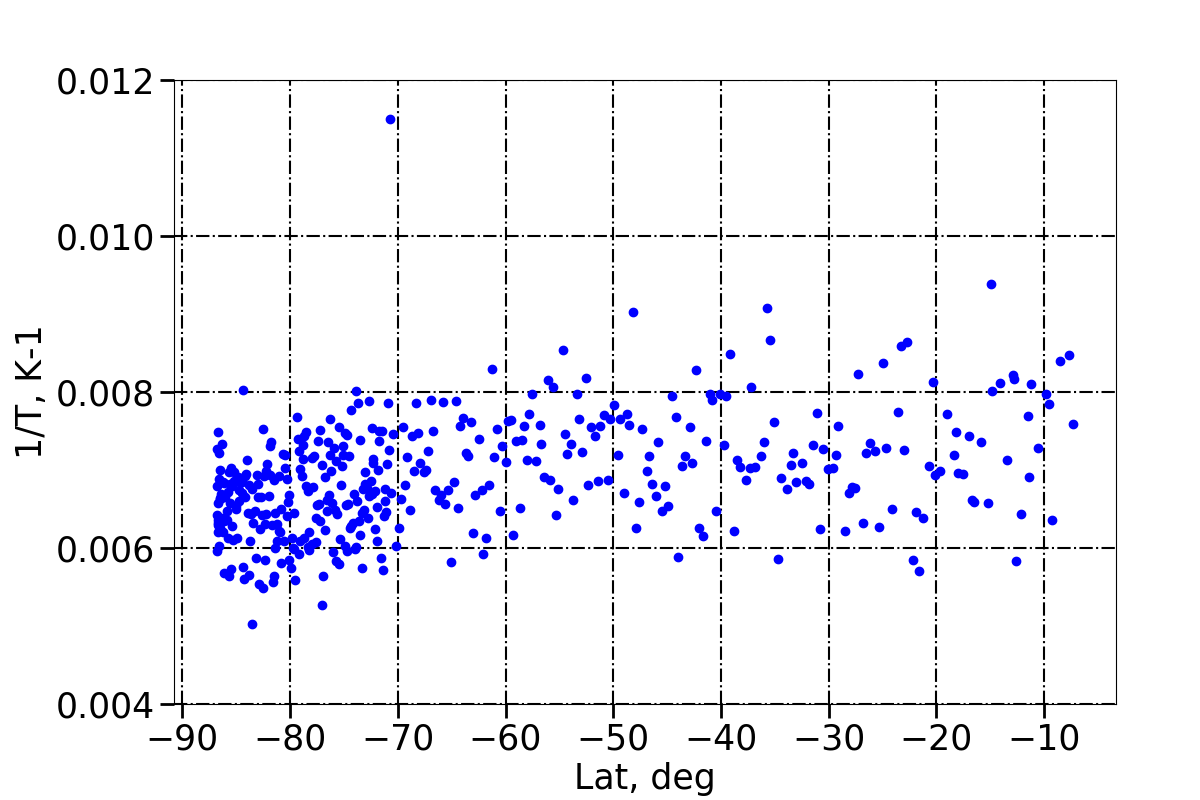}
\end{tabular}
\end{center}
\caption{
From the upper to the lower: MGS, ODY, MRO. From the left to the right: RMS of the relative density calculated over the outbound leg according to the latitude of the orbit's periapsis, inverse of the mean background temperature calculated from the observations over the outbound leg according to the latitude of the orbit's periapsis}
\label{out_reldens_T-1_lat}
\end{figure*}

\begin{figure*}[!tb]
\begin{center}
\begin{tabular}[htb]{ll}
\includegraphics[width=0.5\textwidth]{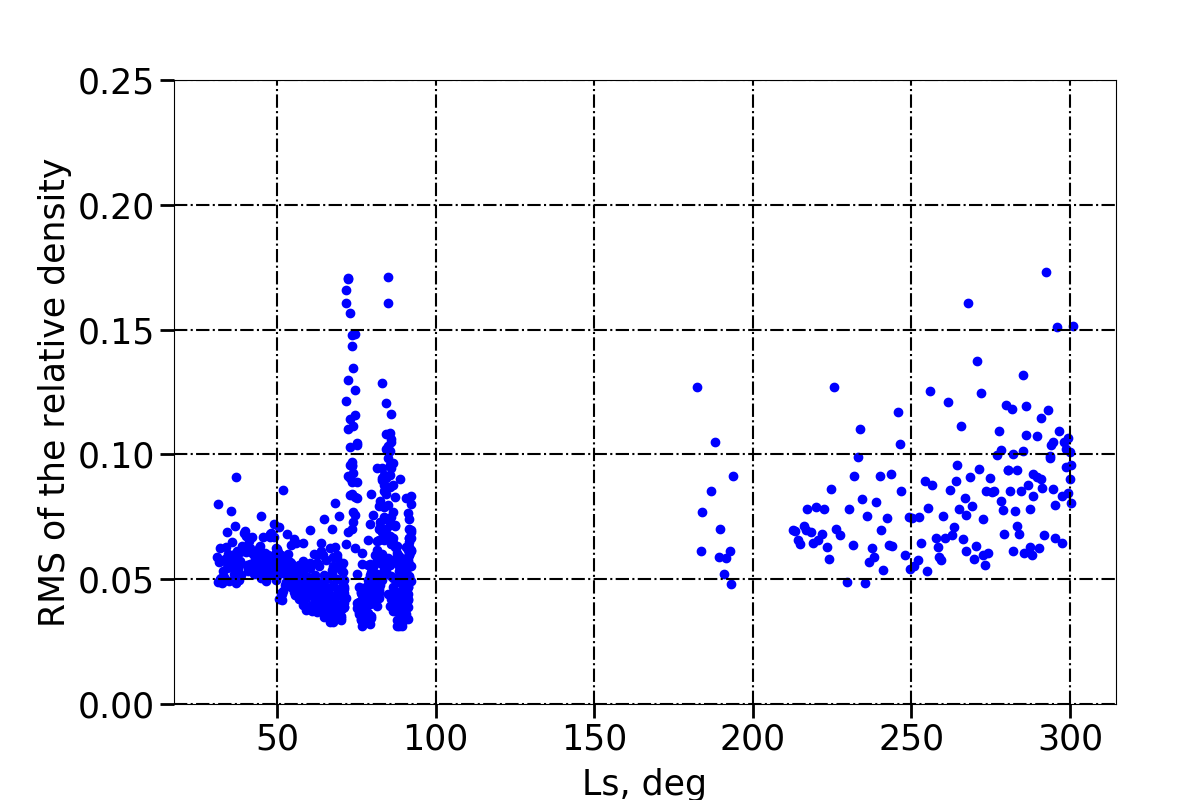}
\includegraphics[width=0.5\textwidth]{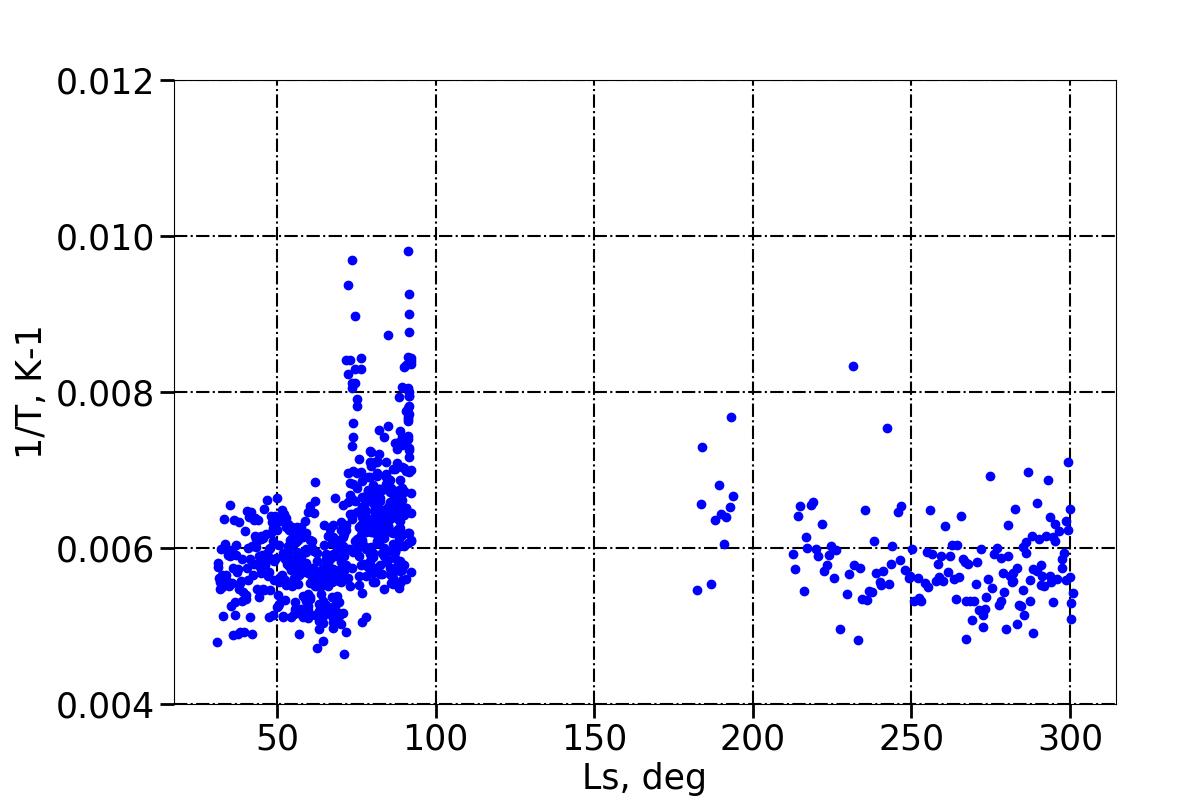}\\
\includegraphics[width=0.5\textwidth]{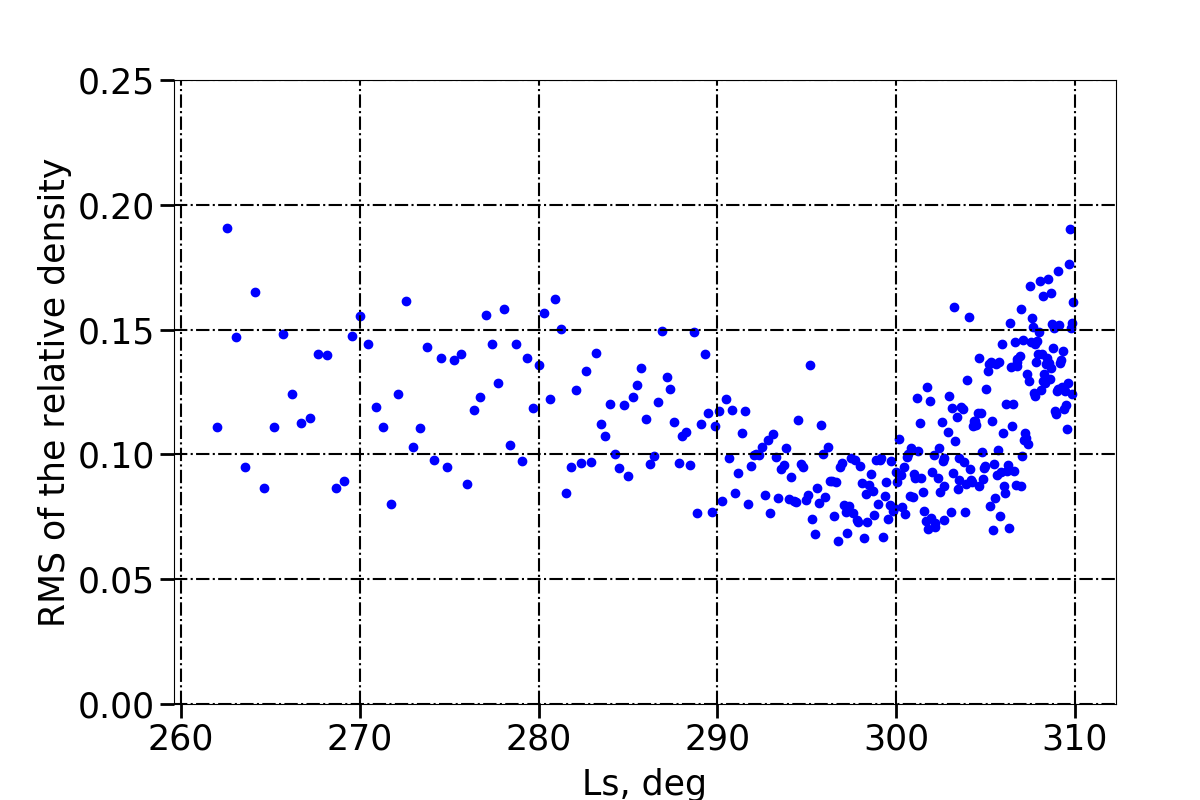}
\includegraphics[width=0.5\textwidth]{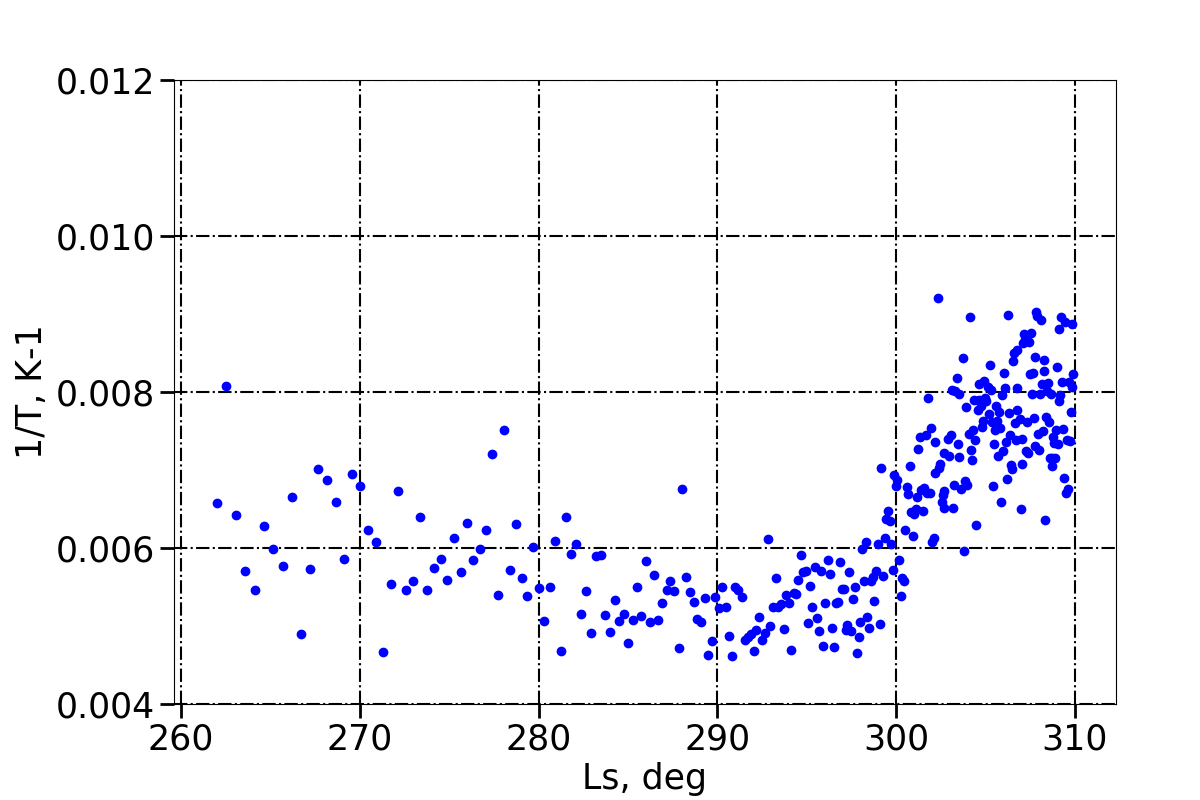}\\
\includegraphics[width=0.5\textwidth]{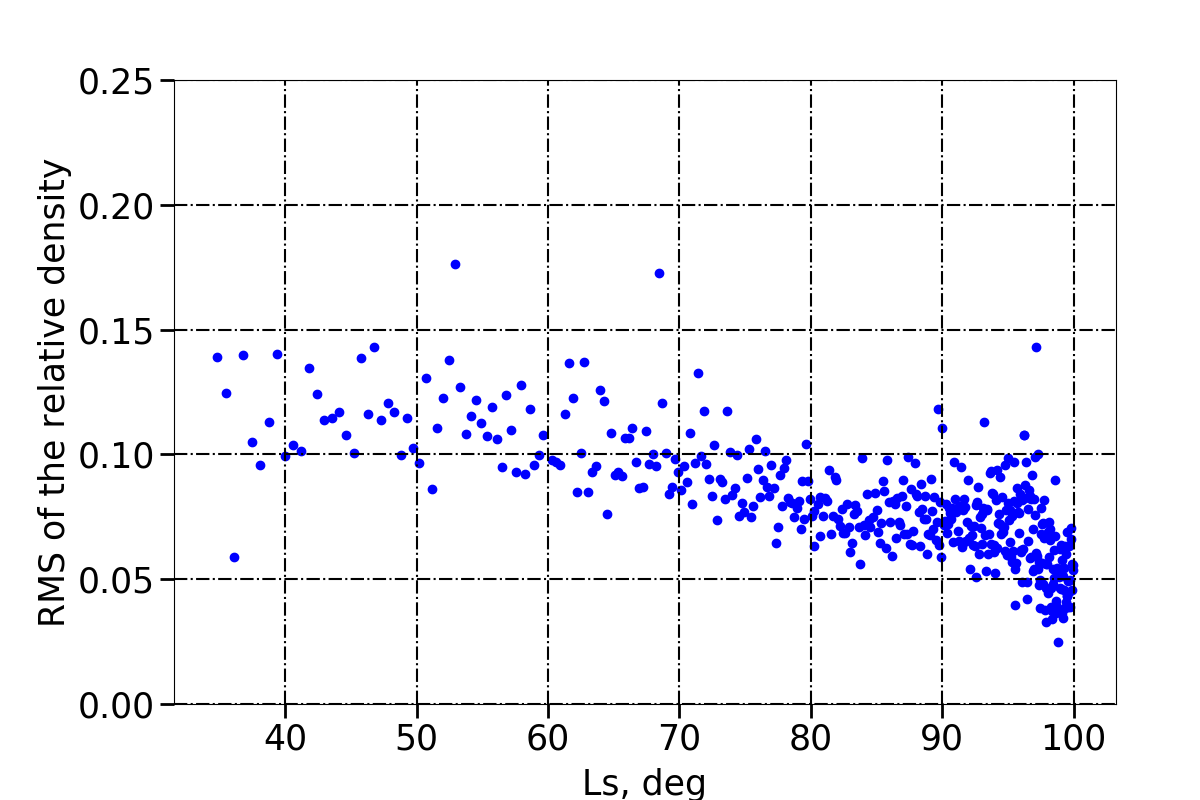}
\includegraphics[width=0.5\textwidth]{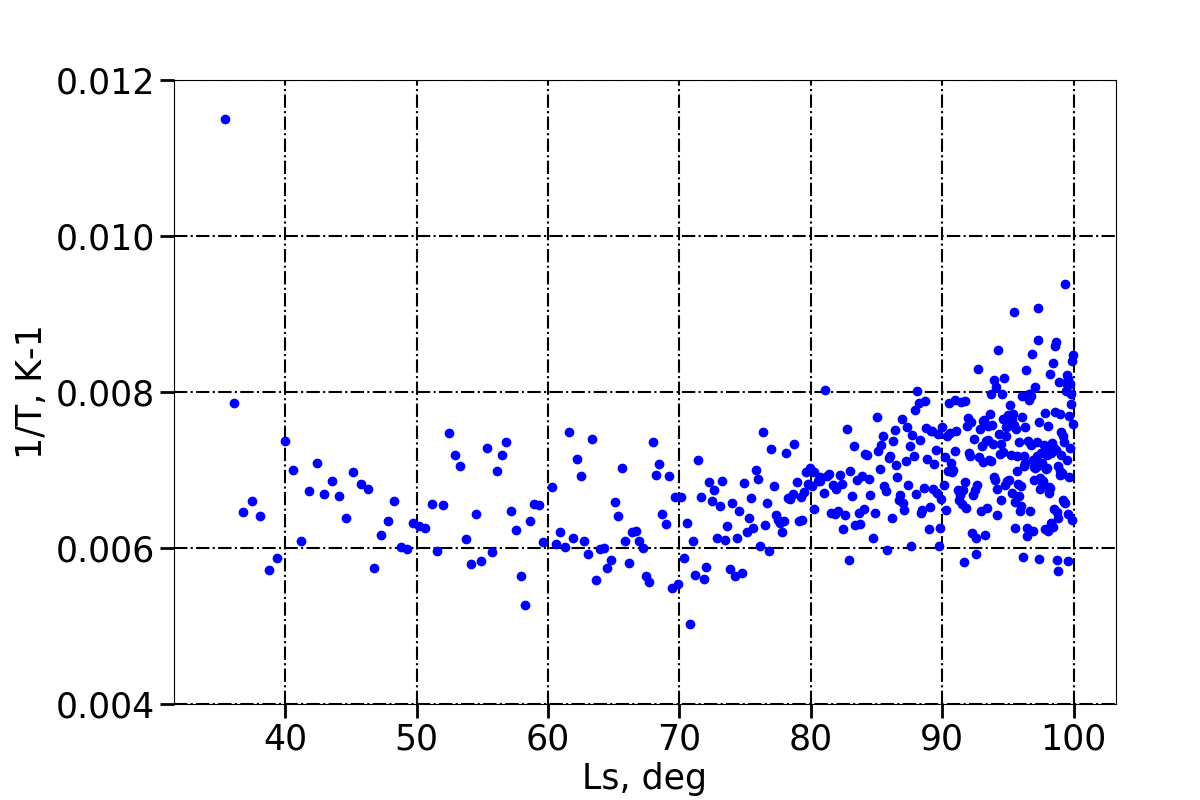}\\
\end{tabular}
\end{center}
\caption{
From the upper to the lower: MGS, ODY, MRO. From the left to the right: RMS of the relative density calculated over the outbound leg according to L$_{s}$, inverse of the mean background temperature calculated from the observations over the outbound leg according to L$_{s}$}
\label{out_reldens_T-1_ls}
\end{figure*}

In Figures \ref{out_reldens_T-1_lat} and  \ref{out_reldens_T-1_ls},  
the observed RMS of the relative density variations 
is compared to
the inverse of the background temperature,
calculated for each orbit of each instrument 
over the outbound leg
(for the sake of brevity, similar results over the inbound leg are not shown).
Latitudinal and seasonal variability are displayed respectively 
in Figures \ref{out_reldens_T-1_lat} and \ref{out_reldens_T-1_ls}.

The amplitude of gravity waves present similar features
with the inverse of temperature in the MGS observations,
with an amplitude increase 
at latitudes 60$^{\circ}$S, 
50$^{\circ}$N 
and particularly at 20$^{\circ}$S,
where inverse temperature is higher in Figure \ref{out_reldens_T-1_lat}, corresponding to L$_s\sim70^{\circ}$ in Figure \ref{out_reldens_T-1_ls}.
The anti-correlation seems easier
to identify in ODY data, 
in particular at polar latitudes around 80$^{\circ}$N, 
where a clear decrease of GWs amplitude is
correlated with the polar warming
(see previous paragraph).
Conversely, no obvious
correlation between density perturbations
and inverse temperature is found in the MRO aerobraking data:
there is an increase in gravity waves activity from latitude -90$^{\circ}$ to -70$^{\circ}$, while the tendency for inverse temperature 
is unclear, corresponding to L$_s\sim35^{\circ}$ in Figure \ref{out_reldens_T-1_ls}. Furthermore, the gravity waves activity decreases at L$_s\sim95^{\circ}$, corresponding to a latitude of -20$^{\circ}$, whereas it is not the case for inverse temperature.
Correlations have been calculated for the three instruments between the gravity waves amplitude and inverse temperature as done for MAVEN in Figure \ref{reldens_T_correlation}, but for all of them the correlation coefficient  $R$ remains below $0.5$. The largest correlation coefficient is obtained for ODY ($R=0.48$), whereas it is around $0.2$ for the two other datasets.



\subsection{Discussion}

The correlation between density perturbations,
caused by gravity waves, and the inverse background temperature,
suggested by equation~\ref{saturation3},
appears to be observed by
MAVEN/NGIMS.
A similar correlation, 
albeit less clear-cut than with the MAVEN/NGIMS dataset, 
is also noticed 
during ODY aerobraking phases at high latitudes.
This correlation seems to be observed at certain locations for MGS, as seen in the previous section, and also in particular cases for MRO, as seen in the previous section for the orbits located in the polar warming. However, for those two spacecrafts, the correlation is not clear at all in the global analysis of the complete datasets.

\begin{figure}[!tb]
\begin{center}
\begin{tabular}[htb]{l}
\includegraphics[width=\textwidth]{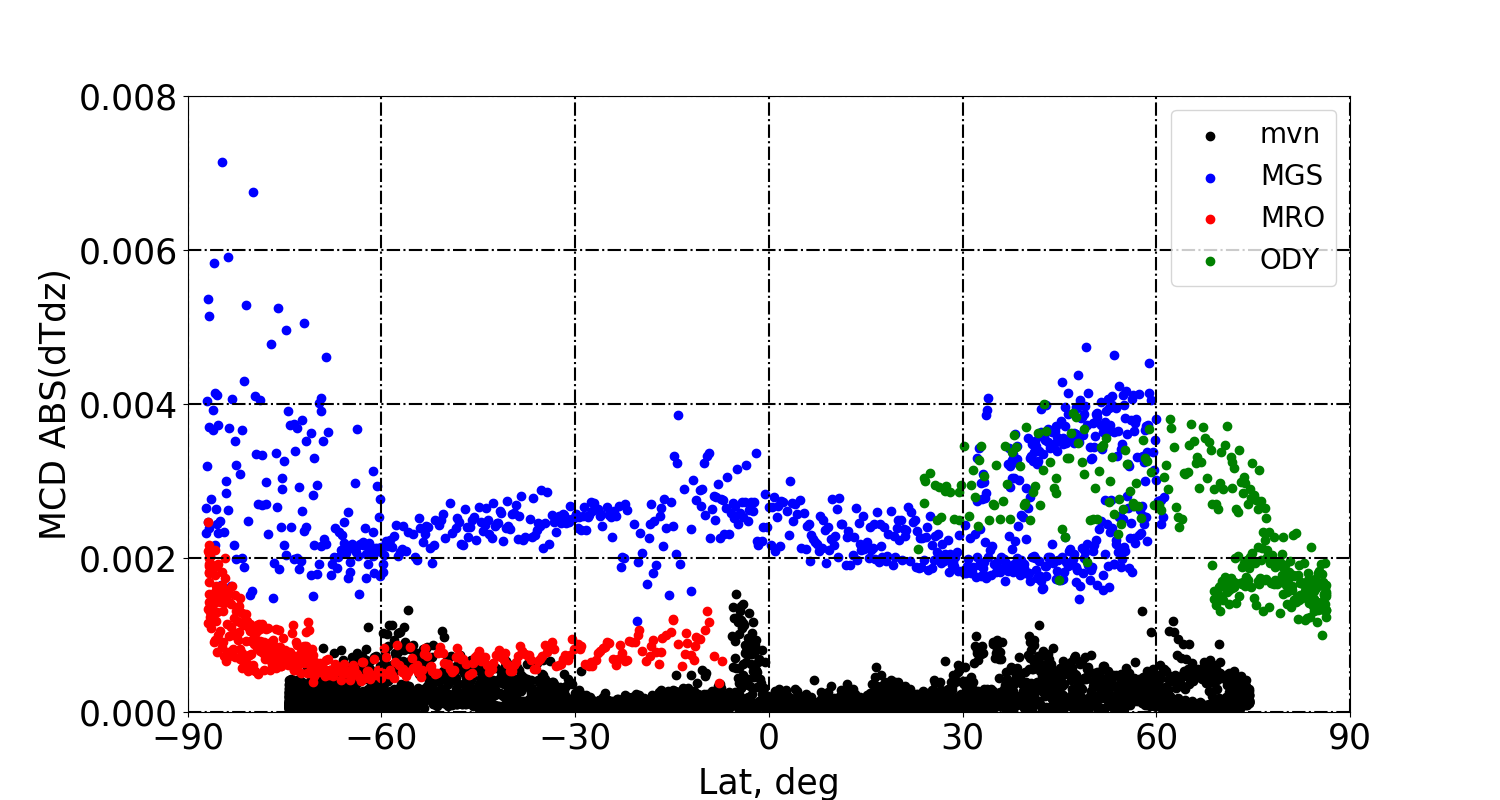}\\
\includegraphics[width=\textwidth]{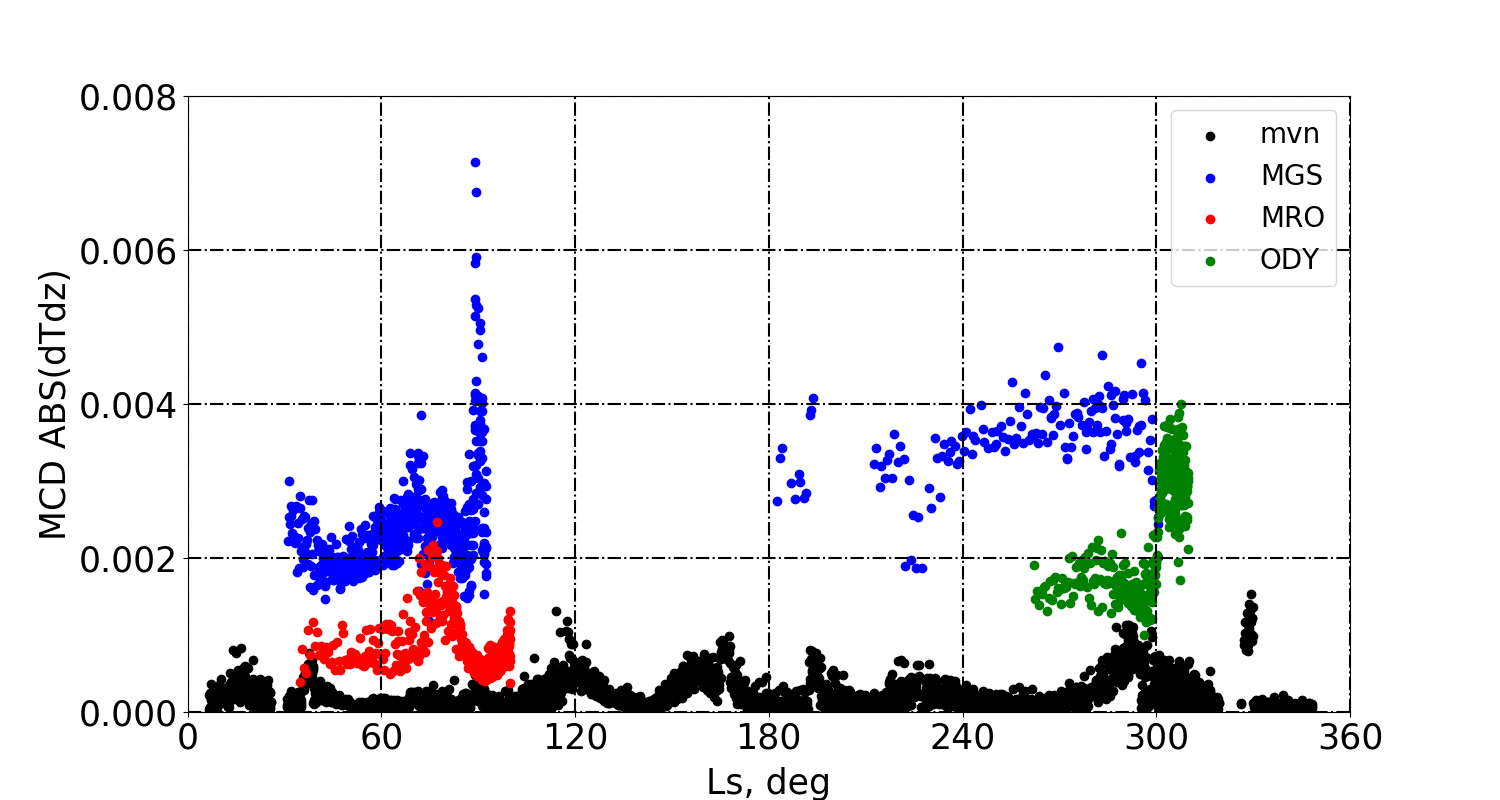}
\end{tabular}
\end{center}
\caption{Upper: Absolute value of the mean temperature gradient along the latitude calculated with the MCD over the outbound leg of each orbit of the different aerobraking instruments and MAVEN/NGIMS. Down: Absolute value of the mean temperature gradient along the L$_{s}$ calculated with the MCD over the outbound leg of each orbit of the different aerobraking instruments and MAVEN/NGIMS.}
\label{dTdz_aero}
\end{figure}

The aerobraking density measurements correspond to 
periapsis conditions at lower altitudes
than the MAVEN/NGIMS measurements 
(cf. Figure~\ref{all_alt_Ls}). 
There, the assumption of isothermal profiles
could not be valid. Indeed, in Figure \ref{dTdz_aero} we compare the temperature gradients calculated with the MCD for the three aerobraking missions along with MAVEN/NGIMS. The Figure shows that MAVEN/NGIMS data mainly correspond to isothermal profiles, whereas the three other instruments present larger temperature gradients. Yet, equation~\ref{saturation3} is only effective in isothermal conditions. As a matter of fact, we observe in the ODY data that a potential correlation between gravity wave activity and inverse temperature only appears where the temperature gradient is the lowest, at higher latitudes. This is also possibly the case for MGS at the points located around latitudes -60$^{\circ}$ and -20$^{\circ}$. However, MRO, which presents the lowest temperature gradient, presents no clear correlation with the inverse temperature. There could be an explanation for the temperature gradient being lower for ODY and MRO, despite the fact they are lower in altitude: at polar latitudes, 
the polar warming shifts the threshold for isothermal
conditions to lower altitudes in the mesosphere.
Furthermore, when the temperature gradient is significant,
the more general equation~\ref{saturation2} 
shall prevail instead of equation~\ref{saturation3},
which means that the amplitude of gravity waves
is proportional to the static stability $N^{2}$
rather than the inverse background temperature.

\begin{figure}[!tb]
\begin{center}
\begin{tabular}[htb]{ll}
\includegraphics[width=6.cm,clip=t]{MGS_out_mreldens_Lat.png}
&
\includegraphics[width=6.cm,clip=t]{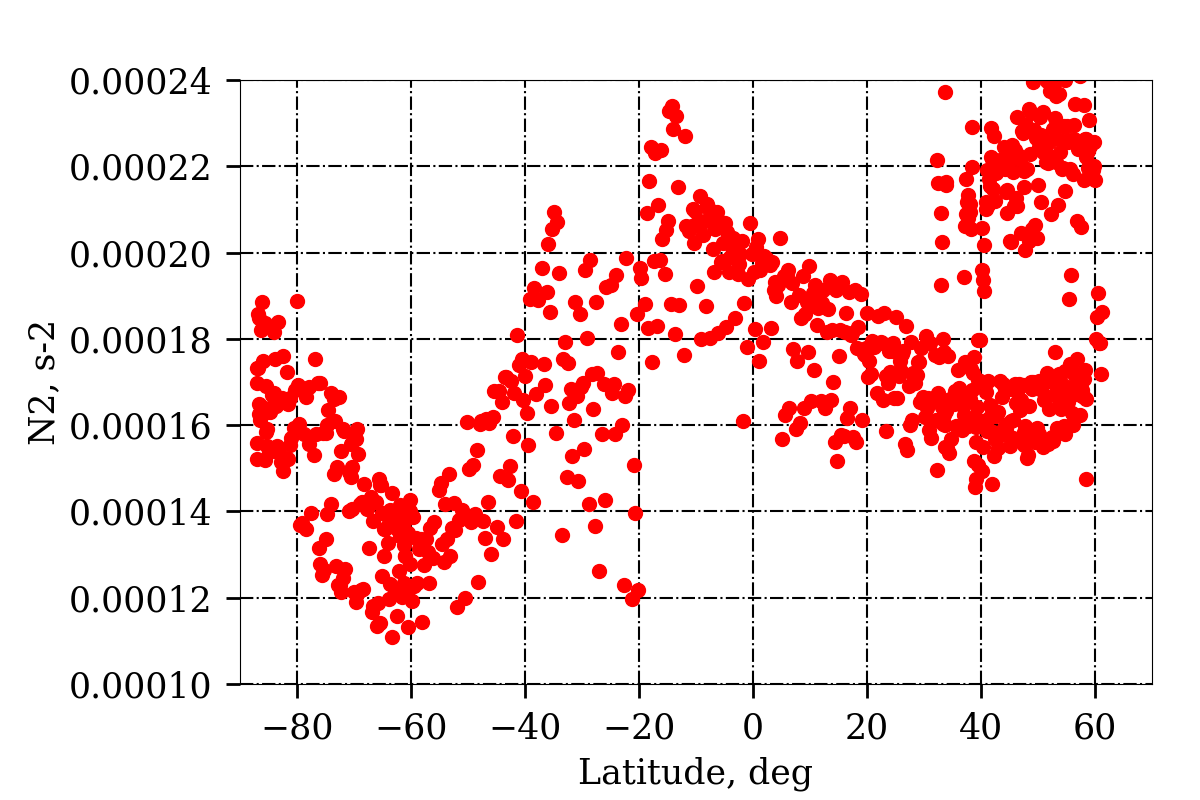}\\
\includegraphics[width=6.cm,clip=t]{ODY_out_mreldens_Lat.png}
&
\includegraphics[width=6.cm,clip=t]{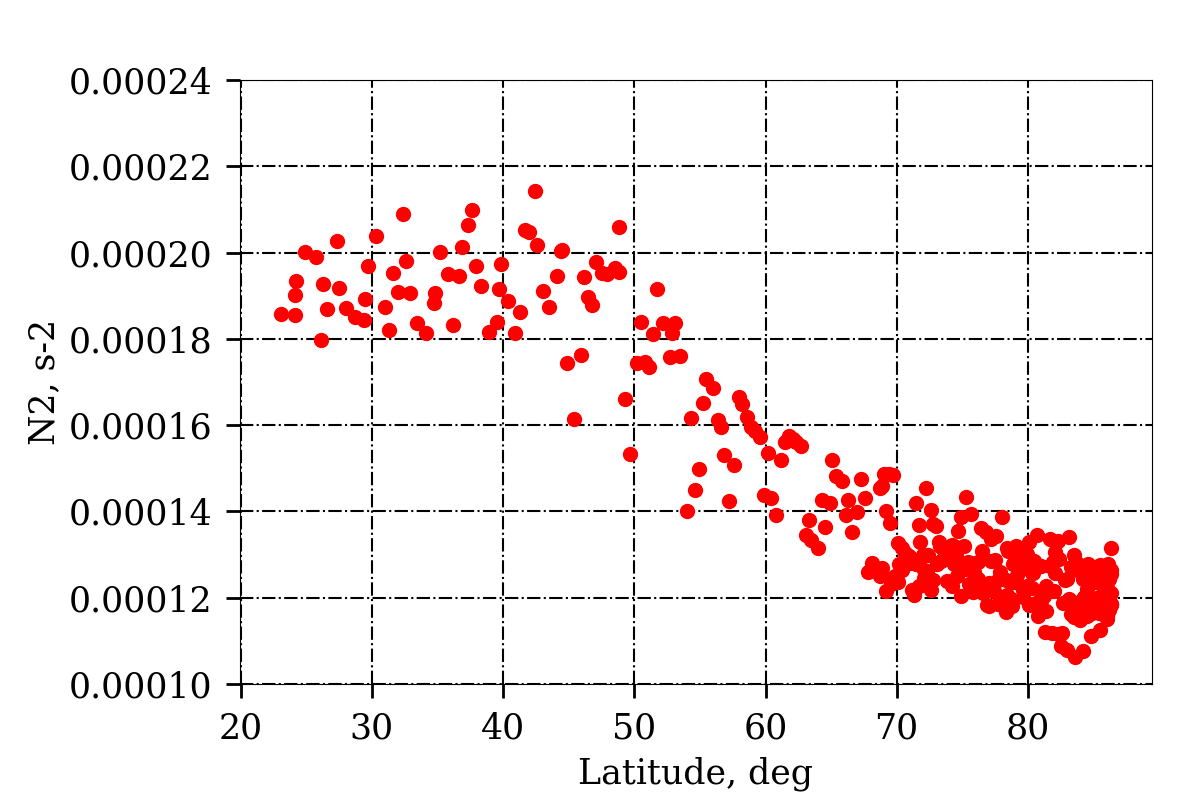}\\
\includegraphics[width=6.cm,clip=t]{MRO_out_mreldens_Lat.png}
&
\includegraphics[width=6.cm,clip=t]{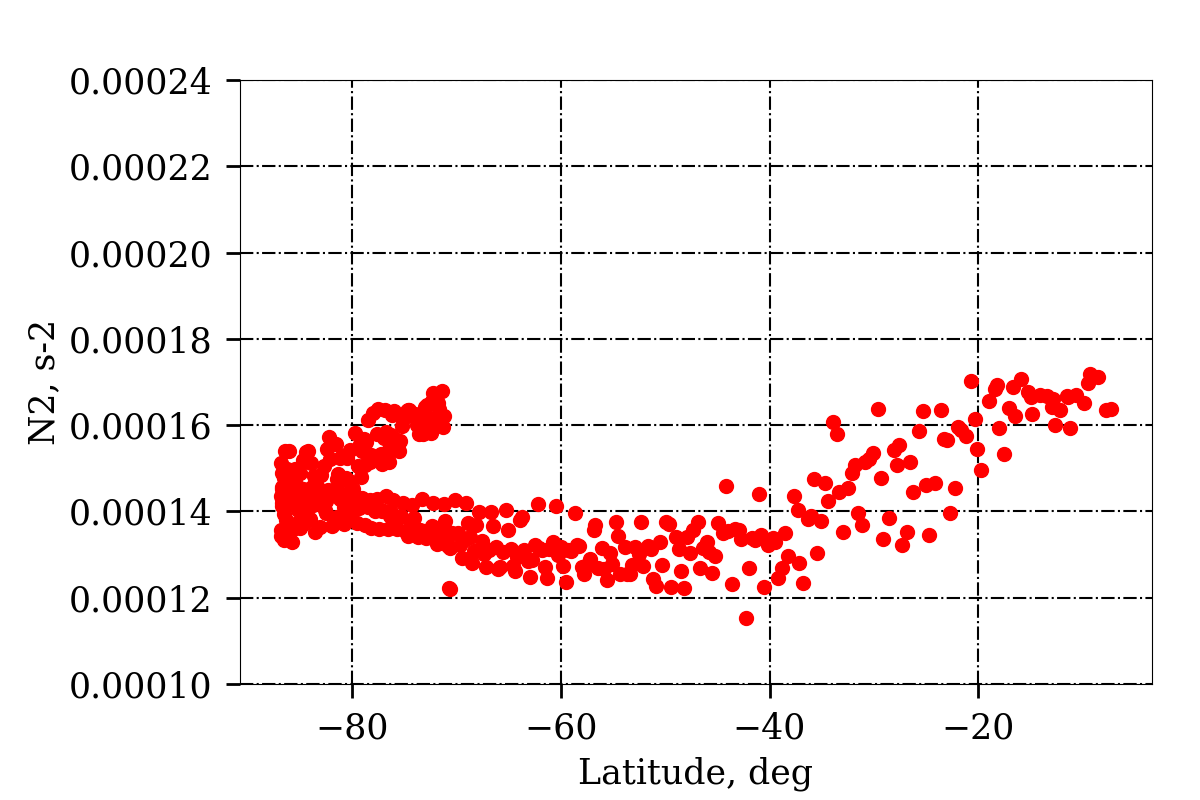}
\end{tabular}
\end{center}
\caption{From the upper to the lower: MGS, ODY, MRO. Left : RMS of the relative density calculated over the outbound leg of each orbit of aerobraking data according to the latitude of the orbit's periapsis ; Right : Mean static stability $N^{2}$ calculated over the outbound leg of each orbit of aerobraking data according to the latitude of the orbit's periapsis, $N^{2}$ has been calculated by means of the Mars Climate Database (MCD) at the different orbital characteristics and with the corresponding dedicated MCD dust scenarios of Mars Year (MY) 25 (MCD detailed document, \citet{Mont:15})}
\label{mro_reldens_N2}
\end{figure}

The possible correlation with static stability~$N^2$ 
can be tested with the MCD
in the conditions that were met by the aerobraking measurements.
Figure \ref{mro_reldens_N2} 
displays the comparison between 
the RMS of the relative density 
acquired at the different aerobraking orbits 
and the static stability calculated from the MCD
(for the corresponding orbital spatio-temporal coordinates).
We observe the same peak of gravity waves activity and static stability for MGS at latitude -20$^{\circ}$ and in the North pole, but not for the other latitudes. A good correlation between the GWs activity and the static stability, as with the inverse temperature, can be found for ODY. Regarding MRO, the static stability~$N^2$ correlates well to 
the observed amplitude of gravity waves
in high-latitude regions
(latitudes above~$-50^{\circ}$S), 
but such a correlation 
is not found at lower latitudes.

There might be multiple reasons for aerobraking measurements not following
equation~\ref{saturation2} in the low and mid latitudes. Firstly, while no correlation was found with potential sources of gravity wave, it is still possible that outside the polar regions, propagation effects would compete with the regional variability of gravity-wave sources. Secondly, following a similar argument, the filtering by critical levels was ruled out for a lack of clear tendency, but might be of peculiar importance for specific regions \citep[see][]{Spig:12gwco2}. Thirdly, the regional variability of vertical wavelength~$k_{z}$, a parameter found in equations~\ref{saturation2} and~\ref{saturation3}, in principle could impact density perturbations \citep{Smit:87}, which then would be less clearly correlated to static stability~$N^2$.

\section{Conclusion \label{sec:conclu}}

We have studied the seasonal and regional variability of density perturbations, putatively caused by the propagation of gravity waves in the thermosphere, in different sets of data issued from the aerobraking phases of MGS, ODY and MRO (accelerometers) and the observations of the NGIMS instrument on board MAVEN. The modeling compiled in the Mars Climate Database has been used to complement background atmospheric conditions obtained by observations. Our conclusions are as follows:
\begin{enumerate}
\item The correlation found in the MAVEN observations by \citet{Tera:17} between the inverse background temperature and the density perturbations reasonably extends to the ODY aerobraking measurements, but not to the MGS and MRO aerobraking measurements.
This result emphasizes the exceptional nature of MAVEN datasets, which combine both isothermal and saturated conditions (equation~\ref{saturation3}).
The seasonal variability of inverse background temperature measured by MAVEN is reproduced in the Mars Climate Database.
\item In comparison to MAVEN/NGIMS measurements, MGS, ODY and MRO aerobraking data cover a lower layer in the thermosphere, where the Mars Climate Database predicts non-isothermal conditions. In these conditions, and under the hypothesis of saturation, a correlation between the gravity waves perturbation with the static stability is expected (equation~\ref{saturation2}).
A correlation of density perturbations monitored both by ODY and MRO during aerobraking in polar regions with static stability~$N^2$ is observed and indicates that wave saturation might be still dominant, but the isothermal conditions are no longer verified (equation~\ref{saturation2}).
\item The spatial variability of gravity-wave-induced density perturbations are difficult to explain for the global MGS dataset and in lower latitudes for ODY and MRO aerobraking, where no clear correlation with neither inverse temperature nor static stability is found. The effects of gravity-wave sources, or wind filtering effects through critical levels, were ruled out as explanations for most of the measured variability, yet might play a stronger role in the low-to-mid latitudes.
\end{enumerate}

Future studies will employ measurements during the aerobraking phase of the ExoMars Trace Gas Orbiter, as well as new measurements by MAVEN, to confirm the conclusions drawn in this study and the existing literature. Broadening the knowledge of gravity wave activity in the mesosphere and thermosphere is crucial to understand the large-scale heat and momentum budget of this part of the Martian atmosphere.

\acknowledgments
The authors acknowledge Centre National d'{\'E}tudes Spatiale (CNES)
and European Space Agency (ESA) for financial support.
We thank two reviewers for thorough and constructive comments
that helped us to improve this paper.






%
%
%
%
%
%
%
%
%
%

\bibliography{newfred}





\listofchanges

\end{document}